\documentclass[twocolumn]{aastex631}

\shorttitle{Slow Wind Connection with Solar Orbiter}
\shortauthors{Yardley et al.}

\begin{document}

\title{Slow Solar Wind Connection Science during Solar Orbiter's First Close Perihelion Passage}

\author[0000-0003-2802-4381]{Stephanie L. Yardley}
\affiliation{Department of Meteorology, University of Reading, Reading, UK}
\affiliation{University College London, Mullard Space Science Laboratory, Holmbury St. Mary, Dorking, Surrey, RH5 6NT, UK}
\affiliation{Donostia International Physics Center (DIPC), Paseo Manuel de Lardizabal 4, 20018 San Sebasti{\'a}n, Spain}

\author[0000-0002-5982-4667]{Christopher J. Owen}
\affiliation{University College London, Mullard Space Science Laboratory, Holmbury St. Mary, Dorking, Surrey, RH5 6NT, UK}

\author[0000-0003-3137-0277]{David M. Long}
\affiliation{University College London, Mullard Space Science Laboratory, Holmbury St. Mary, Dorking, Surrey, RH5 6NT, UK}
\affiliation{Queen's University Belfast, University Road, Belfast BT7 1NN, UK}

\author[0000-0002-0665-2355]{Deborah Baker}
\affiliation{University College London, Mullard Space Science Laboratory, Holmbury St. Mary, Dorking, Surrey, RH5 6NT, UK}

\author[0000-0002-2189-9313]{David H. Brooks}
\affiliation{College of Science, George Mason University, 4400 University Drive, Fairfax, VA 22030 USA}
%\affiliation{Hinode Team, ISAS/JAXA, 3-1-1 Yoshinodai, Chuo-ku, Sagamihara, Kanagawa 252-5210, Japan}

\author[0000-0002-4980-7126]{Vanessa Polito}
\affiliation{Bay Area Environmental Research Institute, NASA Research Park, Moffett Field, CA 94035, USA}
\affiliation{Lockheed Martin Solar \& Astrophysics Laboratory, 3251 Hanover Street, Palo Alto, CA 94304, USA}
\affiliation{Department of Physics, Oregon State University, 301 Weniger Hall, Corvallis, OR 97331}

\author[0000-0002-0665-2355]{Lucie M. Green}
\affiliation{University College London, Mullard Space Science Laboratory, Holmbury St. Mary, Dorking, Surrey, RH5 6NT, UK}

\author[0000-0002-0665-2355]{Sarah Matthews}
\affiliation{University College London, Mullard Space Science Laboratory, Holmbury St. Mary, Dorking, Surrey, RH5 6NT, UK}

\author[0000-0003-2061-2453]{Mathew Owens}
\affiliation{Department of Meteorology, University of Reading, Reading, UK}

\author[0000-0002-7397-2172]{Mike Lockwood}
\affiliation{Department of Meteorology, University of Reading, Reading, UK}

\author[0000-0002-1365-1908]{David Stansby}
\affiliation{Advanced Research Computing Centre, University College London, Gower St, Bloomsbury, London WC1E 6BT, UK}

\author[0000-0001-7927-9291]{Alexander W. James}
\affiliation{University College London, Mullard Space Science Laboratory, Holmbury St. Mary, Dorking, Surrey, RH5 6NT, UK}

\author[0000-0001-7809-0067]{Gherardo Valori}
\affiliation{Max-Planck-Institut f{\"u}r Sonnensystemforschung, Justus-von-Liebig-Weg 3, 37077 G{\"o}ttingen, Germany}

\author{Alessandra Giunta}
\affiliation{RAL Space, UKRI STFC Rutherford Appleton Laboratory, Harwell, Didcot, OX11 0QX, UK}
             
\author[0000-0002-6203-5239]{Miho Janvier}
\affiliation{Universit{\'e} Paris-Saclay, CNRS, Institut d'Astrophysique Spatiale, F-91405, Orsay, France}
\affiliation{Laboratoire Cogitamus, rue Descartes, 75005 Paris, France}

\author[0000-0002-1794-1427]{Nawin Ngampoopun}
\affiliation{University College London, Mullard Space Science Laboratory, Holmbury St. Mary, Dorking, Surrey, RH5 6NT, UK}

\author[0000-0001-8055-0472]{Teodora Mihailescu}
\affiliation{University College London, Mullard Space Science Laboratory, Holmbury St. Mary, Dorking, Surrey, RH5 6NT, UK}

\author[0000-0003-0774-9084]{Andy S. H. To}
\affiliation{University College London, Mullard Space Science Laboratory, Holmbury St. Mary, Dorking, Surrey, RH5 6NT, UK}

\author[0000-0002-2943-5978]{Lidia van Driel-Gesztelyi}
\affiliation{University College London, Mullard Space Science Laboratory, Holmbury St. Mary, Dorking, Surrey, RH5 6NT, UK}
\affiliation{LESIA, Observatoire de Paris, Universit\'e PSL, CNRS, Sorbonne Universit\'e, Univ. Paris Diderot, Sorbonne Paris Cit\'e, 5 place Jules Janssen, 92195 Meudon, France}
\affiliation{Konkoly Observatory, Research Centre for Astronomy and Earth Sciences, Konkoly Thege \'ut 15-17., H-1121, Budapest, Hungary}

\author[0000-0001-8215-6532]{Pascal D\'emoulin}
\affiliation{LESIA, Observatoire de Paris, Universit\'e PSL, CNRS, Sorbonne Universit\'e, Univ. Paris Diderot, Sorbonne Paris Cit\'e, 5 place Jules Janssen, 92195 Meudon, France}
\affiliation{Laboratoire Cogitamus, rue Descartes, 75005 Paris, France}

\author[0000-0003-2647-117X]{Raffaella D'Amicis}
\affiliation{National Institute for Astrophysics, Institute for Space Astrophysics and Planetology, Rome, Italy}

\author[0000-0001-9726-0738]{Ryan J. French}
\affiliation{National Solar Observatory, 3665 Innovation Drive, Boulder CO 80303}

\author[0000-0002-9387-5847]{Gabriel H.H. Suen}
\affiliation{University College London, Mullard Space Science Laboratory, Holmbury St. Mary, Dorking, Surrey, RH5 6NT, UK}

%------------------Magnetic Connectivity------------------
\author[0000-0003-4039-5767]{Alexis P. Roulliard}
\affiliation{IRAP, Universit{\'e} Toulouse III – Paul Sabatier, CNRS, CNES, Toulouse, France}

\author[0000-0001-8247-7168]{Rui F. Pinto}
\affiliation{IRAP, Universit{\'e} Toulouse III – Paul Sabatier, CNRS, CNES, Toulouse, France}
\affiliation{Dept. d’Astrophysique/AIM, CEA/IRFU, CNRS/INSU, Universit\'e Paris-Saclay, 91191 Gif-sur-Yvette Cedex, France}

\author[0000-0002-2916-3837]{Victor R{\'e}ville}
\affiliation{IRAP, Universit{\'e} Toulouse III – Paul Sabatier, CNRS, CNES, Toulouse, France}

%--------------------ESA--------------------------
\author{Christopher J. Watson}
\affiliation{European Space Agency (ESA), European Space Astronomy Centre (ESAC), Camino Bajo del Castillo s/n, E-28692 Villanueva De La Cañada, Madrid, Spain}

\author[0000-0002-1682-1212]{Andrew P. Walsh}
\affiliation{European Space Agency (ESA), European Space Astronomy Centre (ESAC), Camino Bajo del Castillo s/n, E-28692 Villanueva De La Cañada, Madrid, Spain}

\author{Anik De Groof}
\affiliation{European Space Agency (ESA), European Space Astronomy Centre (ESAC), Camino Bajo del Castillo s/n, E-28692 Villanueva De La Cañada, Madrid, Spain}

\author[0000-0001-9922-8117]{David R. Williams}
\affiliation{European Space Agency (ESA), European Space Astronomy Centre (ESAC), Camino Bajo del Castillo s/n, E-28692 Villanueva De La Cañada, Madrid, Spain}

\author[0000-0003-2672-9249]{Ioannis Zouganelis}
\affiliation{European Space Agency (ESA), European Space Astronomy Centre (ESAC), Camino Bajo del Castillo s/n, E-28692 Villanueva De La Cañada, Madrid, Spain}

\author[0000-0001-9027-9954]{Daniel M\"uller}
\affiliation{European Space Agency, ESTEC, Noordwijk, The Netherlands}

%---------------EUI------------------
\author[0000-0003-4052-9462]{David Berghmans}
\affiliation{Solar-Terrestrial Centre of Excellence -- SIDC, Royal Observatory of Belgium, Ringlaan -3, 1180 Brussels, Belgium}

\author[0000-0003-0972-7022]{Fr\'ed\'eric Auch\`ere}
\affiliation{Institut d’Astrophysique Spatiale, CNRS, Univ. Paris-Sud, Universit{\'e} Paris-Saclay, B{\^a}t. 121, 91405}

\author[0000-0001-9457-6200]{Louise Harra}
\affiliation{Physikalisch-Meteorologisches Observatorium Davos, World Radiation Center, 7260, Davos Dorf, Switzerland}
\affiliation{ETH-Z\"urich, Wolfgang-Pauli-Str. 27, 8093 Z\"{u}rich, Switzerland}

\author[0000-0001-6060-9078]{Udo Schuehle}
\affiliation{Max-Planck-Institut f\"ur Sonnensystemforschung, Justus-von-Liebig-Weg 3, 37077 G\"ottingen, Germany}

\author[0000-0001-7090-6180]{Krysztof Barczynski}
\affiliation{Physikalisch-Meteorologisches Observatorium Davos, World Radiation Center, 7260, Davos Dorf, Switzerland}
\affiliation{ETH-Z\"urich, Wolfgang-Pauli-Str. 27, 8093 Z\"urich, Switzerland}

\author[0000-0003-4290-1897]{\'Eric Buchlin}
\affiliation{Institut d’Astrophysique Spatiale, CNRS, Univ. Paris-Sud, Universit{\'e} Paris-Saclay, B{\^a}t. 121, 91405}

%\author{Pradeep Chitta}
%\affiliation{Max-Planck-Institut f\"ur Sonnensystemforschung, Justus-von-Liebig-Weg 3, 37077 G\"ottingen, Germany}

\author[0000-0003-1294-1257]{Regina Aznar Cuadrado}
\affiliation{Max-Planck-Institut f{\"u}r Sonnensystemforschung, Justus-von-Liebig-Weg 3, 37077 G{\"o}ttingen, Germany}

\author[0000-0002-2265-1803]{Emil Kraaikamp}
\affiliation{Solar-Terrestrial Centre of Excellence -- SIDC, Royal Observatory of Belgium, Ringlaan -3, 1180 Brussels, Belgium}

\author{Sudip Mandal}
\affiliation{Max-Planck-Institut f{\"u}r Sonnensystemforschung, Justus-von-Liebig-Weg 3, 37077 G{\"o}ttingen, Germany}

\author[0000-0003-1438-1310]{Susanna Parenti}
\affiliation{Institut d’Astrophysique Spatiale, CNRS, Univ. Paris-Sud, Universit{\'e} Paris-Saclay, B{\^a}t. 121, 91405}

\author[0000-0001-9921-0937]{Hardi Peter}
\affiliation{Max-Planck-Institut f{\"u}r Sonnensystemforschung, Justus-von-Liebig-Weg 3, 37077 G{\"o}ttingen, Germany}

\author[0000-0002-6097-374X]{Luciano Rodriguez}
\affiliation{Solar-Terrestrial Centre of Excellence -- SIDC, Royal Observatory of Belgium, Ringlaan -3, 1180 Brussels, Belgium}

\author[0000-0002-7669-5078]{Conrad Schwanitz}
\affiliation{Physikalisch-Meteorologisches Observatorium Davos, World Radiation Center, 7260, Davos Dorf, Switzerland}
\affiliation{ETH-Z{\"u}rich, Wolfgang-Pauli-Str. 27, 8093 Z{\"u}rich, Switzerland}

\author{Phil Smith}
\affiliation{University College London, Mullard Space Science Laboratory, Holmbury St. Mary, Dorking, Surrey, RH5 6NT, UK}

\author[0000-0001-7298-2320]{Luca Teriaca}
\affiliation{Max-Planck-Institut f{\"u}r Sonnensystemforschung, Justus-von-Liebig-Weg 3, 37077 G{\"o}ttingen, Germany}

\author[0000-0002-5022-4534]{Cis Verbeeck}
\affiliation{Solar-Terrestrial Centre of Excellence -- SIDC, Royal Observatory of Belgium, Ringlaan -3, 1180 Brussels, Belgium}

\author[0000-0002-2542-9810]{Andrei N. Zhukov}
\affiliation{Solar-Terrestrial Centre of Excellence -- SIDC, Royal Observatory of Belgium, Ringlaan -3, 1180 Brussels, Belgium}
\affiliation{Skobeltsyn Institute of Nuclear Physics, Moscow State University, 119992 Moscow, Russia}

%---------------IRIS---------------

\author[0000-0002-8370-952X]{Bart De Pontieu}
\affiliation{Lockheed Martin Solar \& Astrophysics Laboratory, 3251 Hanover Street, Palo Alto, CA 94304, USA}
\affiliation{Rosseland Centre for Solar Physics, University of Oslo, P.O. Box 1029 Blindern, NO-0315 Oslo, Norway}
\affiliation{Institute of Theoretical Astrophysics, University of Oslo, P.O. Box 1029 Blindern, NO-0315 Oslo, Norway}

%---------------MAG----------------
\author[0000-0002-7572-4690]{Tim Horbury}
\affiliation{The Blackett Laboratory, Imperial College London, London, SW7 2AZ, UK}

%---------------PHI----------------
\author[0000-0002-3418-8449]{Sami K. Solanki}
\affiliation{Max-Planck-Institut f{\"u}r Sonnensystemforschung, Justus-von-Liebig-Weg 3, 37077 G{\"o}ttingen, Germany}

\author[0000-0002-3387-026X]{Jose Carlos del Toro Iniesta}
\affiliation{Instituto de Astrof{\'i}sica de Andaluc{\'i}a (IAA-CSIC), Apartado de Correos 3004, E-18080 Granada, Spain}
 
\author{Joachim Woch}
\affiliation{Max-Planck-Institut f{\"u}r Sonnensystemforschung, Justus-von-Liebig-Weg 3, 37077 G{\"o}ttingen, Germany}

\author[0000-0002-9972-9840]{Achim Gandorfer}
\affiliation{Max-Planck-Institut f{\"u}r Sonnensystemforschung, Justus-von-Liebig-Weg 3, 37077 G{\"o}ttingen, Germany}

\author{Johann Hirzberger}
\affiliation{Max-Planck-Institut f{\"u}r Sonnensystemforschung, Justus-von-Liebig-Weg 3, 37077 G{\"o}ttingen, Germany}

\author[0000-0001-8829-1938]{David Orozco S{\'u}arez}
\affiliation{Instituto de Astrof{\'i}sica de Andaluc{\'i}a (IAA-CSIC), Apartado de Correos 3004, E-18080 Granada, Spain}

\author[0000-0002-1790-1951]{Thierry Appourchaux}
\affiliation{Univ. Paris-Sud, Institut d’Astrophysique Spatiale, UMR 8617, CNRS, B\^ atiment 121, 91405 Orsay Cedex, France}

\author[0000-0003-2755-5295]{Daniele Calchetti}
\affiliation{Max-Planck-Institut f{\"u}r Sonnensystemforschung, Justus-von-Liebig-Weg 3, 37077 G{\"o}ttingen, Germany}

\author[0000-0002-5387-636X]{Jonas Sinjan}
\affiliation{Max-Planck-Institut f{\"u}r Sonnensystemforschung, Justus-von-Liebig-Weg 3, 37077 G{\"o}ttingen, Germany}

\author[0000-0002-4796-9527]{Fatima Kahil}
\affiliation{Max-Planck-Institut f{\"u}r Sonnensystemforschung, Justus-von-Liebig-Weg 3, 37077 G{\"o}ttingen, Germany}

\author[0000-0002-3776-9548]{Kinga Albert}
\affiliation{Max-Planck-Institut f{\"u}r Sonnensystemforschung, Justus-von-Liebig-Weg 3, 37077 G{\"o}ttingen, Germany}

\author{Reiner Volkmer}
\affiliation{Leibniz-Institut für Sonnenphysik, Sch\"oneckstr. 6, D-79104 Freiburg, Germany}

%--------------SPICE---------------

\author[0000-0001-9218-3139]{Mats Carlsson}
\affiliation{Institute of Theoretical Astrophysics, University of Oslo, Oslo, Norway}

\author[0000-0002-6093-7861]{Andrzej Fludra}
\affiliation{RAL Space, UKRI STFC Rutherford Appleton Laboratory, Didcot, Oxfordshire, OX11 0QX, UK}

\author[0000-0001-8830-1200]{Don Hassler}
\affiliation{Southwest Research Institute, Boulder, CO 80302, USA}

\author{Martin Caldwell}
\affiliation{RAL Space, UKRI STFC Rutherford Appleton Laboratory, Didcot, Oxfordshire, OX11 0QX, UK}

\author[0000-0002-8673-3920]{Terje Fredvik}
\affiliation{Institute of Theoretical Astrophysics, University of Oslo, Oslo, Norway}

\author{Tim Grundy}
\affiliation{RAL Space, UKRI STFC Rutherford Appleton Laboratory, Didcot, Oxfordshire, OX11 0QX, UK}

\author{Steve Guest}
\affiliation{RAL Space, UKRI STFC Rutherford Appleton Laboratory, Didcot, Oxfordshire, OX11 0QX, UK}

\author[0000-0001-8007-9764]{Margit Haberreiter}
\affiliation{Physikalisch-Meteorologisches Observatorium Davos, World Radiation Center, 7260, Davos Dorf, Switzerland}

\author{Sarah Leeks}
\affiliation{RAL Space, UKRI STFC Rutherford Appleton Laboratory, Didcot, Oxfordshire, OX11 0QX, UK}

\author[0000-0002-0397-2214]{Gabriel Pelouze}
\affiliation{Universit{\'e} Paris-Saclay, CNRS, Institut d'Astrophysique Spatiale, F-91405, Orsay, France}

\author[0000-0001-7016-7226]{Joseph Plowman}
\affiliation{Southwest Research Institute, Boulder, CO 80302, USA}

\author[0000-0003-1159-5639]{Werner Schmutz}
\affiliation{Physikalisch-Meteorologisches Observatorium Davos, World Radiation Center, 7260, Davos Dorf, Switzerland}

\author{Sunil Sidher}
\affiliation{RAL Space, UKRI STFC Rutherford Appleton Laboratory, Didcot, Oxfordshire, OX11 0QX, UK}

%\author[0000-0001-7298-2320]{Luca Teriaca}
%\affiliation{Max-Planck-Institut f{\"u}r Sonnensystemforschung, Justus-von-Liebig-Weg 3, 37077 G{\"o}ttingen, Germany}

\author[0000-0002-6895-6426]{William T. Thompson}
\affiliation{ADNET Systems Inc., NASA Goddard Space Flight Center, Greenbelt, MD 20771, USA}

%\author[0000-0001-9034-2925]{Peter R. Young}
%\affiliation{NASA Goddard Space Flight Center, Greenbelt, MD 20771, USA}
%\affiliation{Department of Mathematics, Physics and Electrical Engineering, Northumbria University, Newcastle upon Tyne, NE7 7XA, UK}

%---------------SWA----------------
\author[0000-0003-2783-0808]{Philippe Louarn}
\affiliation{Institut de Recherche en Astrophysique et Planétologie, CNRS, Université de Toulouse, CNES, Toulouse, France}

\author[0000-0002-9975-0148]{Andrei Federov}
\affiliation{Institut de Recherche en Astrophysique et Planétologie, CNRS, Université de Toulouse, CNES, Toulouse, France}

%-----------------------------------------------

\correspondingauthor{Stephanie L. Yardley}
\email{s.l.yardley@reading.ac.uk}

\begin{abstract}
The Slow Solar Wind Connection Solar Orbiter Observing Plan (Slow Wind SOOP) was developed to utilise the extensive suite of remote sensing and in situ instruments on board the ESA/NASA Solar Orbiter mission to answer significant outstanding questions regarding the origin and formation of the slow solar wind. The Slow Wind SOOP was designed to link remote sensing and in situ measurements of slow wind originating at open–closed magnetic field boundaries. The SOOP ran just prior to Solar Orbiter’s first close perihelion passage during two remote sensing windows (RSW1 and RSW2) between 2022 March 3--6 and 2022 March 17--22, while Solar Orbiter was at a heliocentric distance of 0.55--0.51 and 0.38--0.34~au from the Sun, respectively. Coordinated observation campaigns were also conducted by Hinode and IRIS. The magnetic connectivity tool was used, along with low latency in situ data, and full-disk remote sensing observations, to guide the target pointing of Solar Orbiter. Solar Orbiter targeted an active region complex during RSW1, the boundary of a coronal hole, and the periphery of a decayed active region during RSW2. Post-observation analysis using the magnetic connectivity tool along with in situ measurements from MAG and SWA/PAS, show that slow solar wind, with velocities between $\sim$210 and 600~km~s$^{-1}$, arrived at the spacecraft originating from two out of three of the target regions. The Slow Wind SOOP, despite presenting many challenges, was very successful, providing a blueprint for planning future observation campaigns that rely on the magnetic connectivity of Solar Orbiter.
 
\end{abstract}

\keywords{Solar wind (1534); Slow solar wind (1873); Solar active regions (1974); Solar coronal holes (1484) }

\section{Introduction} \label{sec:intro}

%{\it Solar Wind Intro}\\
The solar wind is the continuous stream of hot, tenuous plasma that flows away from the Sun, carrying magnetic flux into the heliosphere. Historical observations have shown that the solar wind generally exists in two relatively different regimes: slow ($\lesssim$~500~km~s$^{-1}$) and fast ($>$~500 km~s$^{-1}$) wind streams. However, the distinction between these two different solar wind streams should not be made based on speed alone.

Despite several decades of remote sensing observations and in situ measurements made by a multitude of spacecraft, there is still no consensus on the origins of the slow solar wind. It is well known that the source of the fast solar wind is open magnetic field that is rooted in dark regions of the solar atmosphere seen in EUV known as coronal holes (CH; \citealt{wilcox1968, krieger1973, zirker1977, cranmer2009}). However, there are still many open questions regarding the formation of the slow solar wind including where does the slow solar wind originate, and how is it heated, released, and accelerated into the heliosphere? For recent reviews on this topic see \citealt{abbo2016, cranmer2017, viall2020}.

While the slow solar wind is generally characterised by properties such as low proton speeds, high proton temperatures, dense plasma, large mass flux, greater variability, and high values of first ionisation potential (FIP) bias \citep{abbo2016}, slow wind can at times exhibit properties similar to the fast wind including high Alfv{\'e}nicity (high-degree of correlation between magnetic field and velocity components) or a low FIP bias. These varying properties suggest that slow wind likely originates from a multitude of source regions with differing properties, multiple release, and acceleration mechanisms.

%{\it AR outflows}\\
One promising candidate that has recently been proposed as a source of the slow solar wind are upflows at the boundaries of active regions (ARs). Spectroscopic observations taken by the Extreme ultraviolet Imaging Spectrometer (EIS; \citealt{Culhane2007}) on board the Hinode spacecraft \citep{kosugi2007} have revealed persistent blue shifts in coronal lines such as Fe XII 195.12~\AA~indicating the presence of upflowing plasma (\citealt{sakao2007, harra2008, baker2009, brooks2015, hinode2019, stansby2021, tian2021, yardley2021, brooks2022} and references therein). These upflows are present during the entire lifetime of active regions (e.g. \citealt{demoulin2013, baker2017,brooks2021a}). If these upflows are associated with open field lines, then this plasma is able to escape outwards and into the heliosphere as the solar wind. Remote sensing and in situ abundance measurements have also demonstrated the contribution of plasma originating from active regions to the solar wind \citep{harra2008, wang2010, brooks2011, vandriel2012, brooks2015, macneil2019, stansby2020}.

Not all active regions that exhibit upflows are associated with open magnetic field. In these cases, upflows can become outflows if the active region is in close proximity to a coronal hole \citep{vandriel2012, edwards2016}. Alternatively, upflowing plasma could escape indirectly through interchange reconnection between the closed magnetic field loops of an AR and adjacent open magnetic field \citep{crooker2012, mandrini2014, owens2020, brooks2021b}. 

%{\it Coronal hole boundaries}
Another source that can also contribute to the slow solar wind are coronal holes (CHs), although these are mainly responsible for fast solar wind streams. Two ways that coronal holes can contribute is through interchange reconnection between the CH boundary and neighbouring magnetic flux \citep{fisk2001, antiochos2011} or through rapidly diverging magnetic field present inside small CHs, or at the boundary of large CHs \citep{cranmer2009}. The latter is based on the inverse relationship between the solar wind speed at 1~au and expansion rate of magnetic flux tubes, typically determined at the source surface height of 2.5~R$_{\odot}$ using coronal magnetic field extrapolations (e.g. \citealt{wang1990}). Recent analysis by \citet{stansby2019} found that slow wind with high Alfv{\'e}nicity detected between 0.3 and 0.4~au supports this scenario, as the high Alfv{\'e}nicity could indicate outflows along open magnetic field. Alfv{\'e}nic slow wind has also been detected at 1~au originating from CH boundaries or small equatorial CHs \citep{damicis2015, damicis2019,wang2019}. Finally, another contributor to the slow solar wind are small blobs that are visible in coronagraph observations and are continuously released and travel outwards from the cusps of helmet streamers, known as streamer blobs (e.g. \citealt{sheeley2009, rouillard2010a, rouillard2010b, rouillard2011}).

%{\it Solar Orbiter}\\
In order to provide a complete picture of the solar wind including its origin, heating and acceleration mechanisms, transport processes, and composition, the joint ESA/NASA Solar Orbiter mission \citep{muller2020,garcia2021} was launched from Cape Canaveral, Florida on 10 February 2020. Solar Orbiter hosts a payload consisting of six remote sensing \citep{auchere2020} and four in situ instruments \citep{walsh2020}, which are now working together to answer open questions regarding the origin and formation of the slow solar wind by connecting slow solar wind measured by in situ instruments with remote sensing observations of its source regions. Given the limited spatial field of view and temporal observing windows due to its unique orbit, remote sensing observations taken by Solar Orbiter need to be planned and coordinated meticulously in advance. To achieve Solar Orbiter's science goals, Solar Orbiter Observing Plans (SOOPs) have been designed. For details regarding Solar Orbiter's Science Activity Plan, mission planning and the individual SOOPs refer to \citet{zouganelis2020}. The coordination of the science operations for the remote-sensing and in situ instruments are governed by the Remote Sensing Working Group \citep{auchere2020} and the In situ Working Group \citep{walsh2020}.

%{\it Slow Wind SOOP}\\
In this paper, we describe the planning, observations and preliminary results of the Slow Solar Wind SOOP that operated during Solar Orbiter's first perihelion in March 2022.

\section{Slow Solar Wind Connection Solar Orbiter Observing Plan}

\begin{figure}[t!]
\centering
\includegraphics[width=0.48\textwidth]{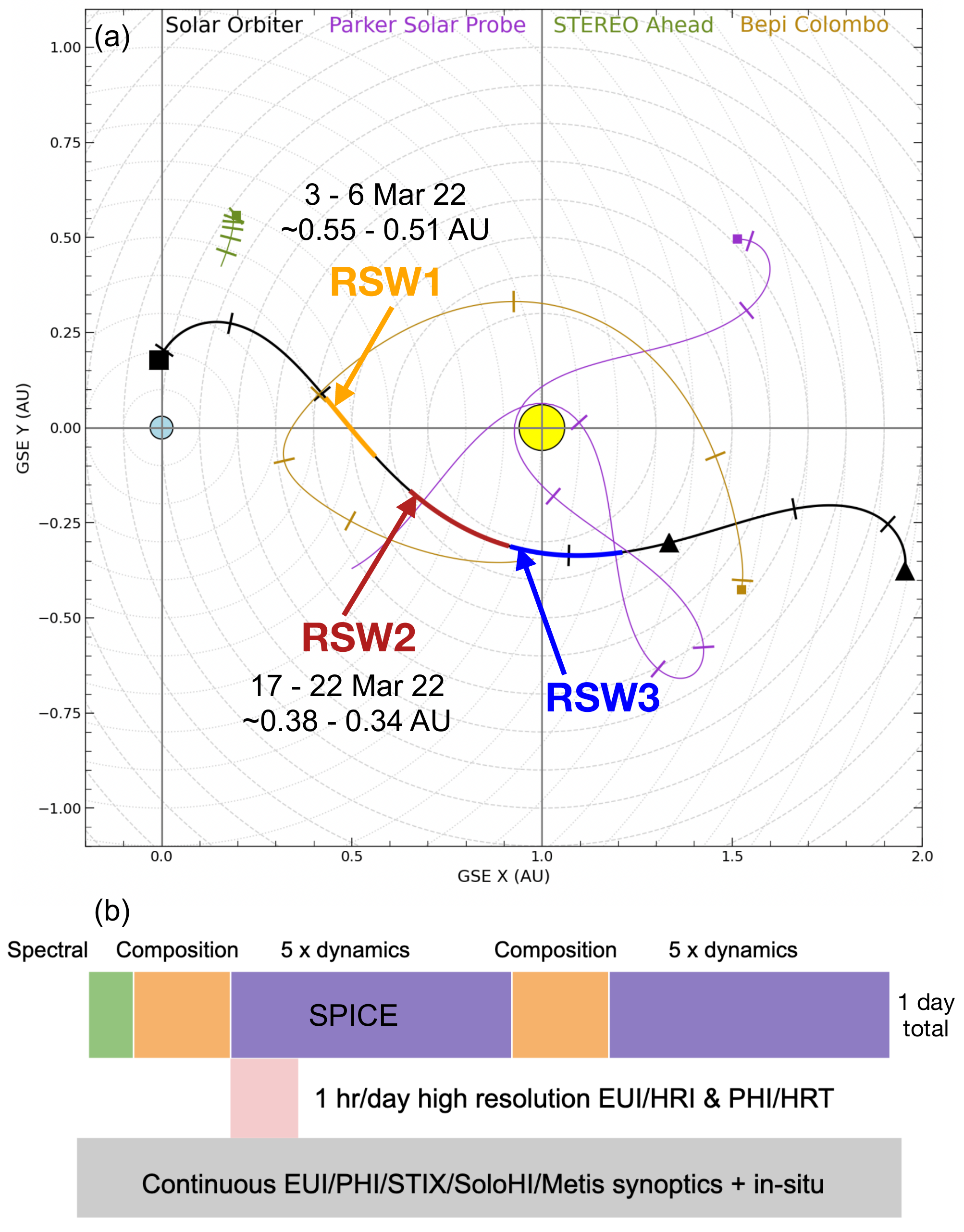}
\caption{Panel~a shows the projection of Solar Orbiter's orbit in the ecliptic plane in geocentric solar ecliptic (GSE) coordinates, during long term planning period number 6 (LTP6; from 2021 December 27 until 2022 April 11). This time period includes the three remote sensing windows (yellow, red, blue) during the first perihelion of Solar Orbiter. RSW1 and RSW2 refer to time period of the two remote-sensing windows of the slow wind SOOP in March 2022. Panel~b shows the remote sensing observations and in situ measurements designed as part of the Slow Wind SOOP. \label{fig:fig1}}
\end{figure}

The Slow Wind SOOP or L\_SMALL\_HRES\_HCAD\_Slow-Wind-Connection (Section 4.14 of \citealt{zouganelis2020}) has been designed to identify the sources of the slow solar wind and address the top-level science goal of Solar Orbiter: ``What drives the solar wind and where does the coronal magnetic field originate from?" More specifically, the SOOP aims to:

\begin{itemize}
    \item link plasma observed in situ to specific coronal source regions,
    \item observe the magnetic field configuration and its evolution at open-closed boundaries,
    \item identify possible signatures of interchange reconnection at open-closed boundaries.
\end{itemize}

The primary targets of the Slow Wind SOOP are therefore either the periphery of an active region or the boundary of a coronal hole.

While the in situ instruments on board Solar Orbiter can take measurements almost continuously, the remote sensing instruments take full disk synoptics throughout the orbit but high-resolution observations only during three $\sim$10-day windows, that during low-inclination orbits are typically centered around the closest approach to the Sun. The Slow Wind SOOP operated at the beginning of the first and second remote-sensing windows (RSWs 1 and 2), prior to Solar Orbiter's first close perihelion passage on 2022 March 26 (see Figure~\ref{fig:fig1}~a, adapted from the Solar Orbiter website\footnote{\url{https://issues.cosmos.esa.int/solarorbiterwiki/display/
SOSP/LTP06+Q1-2022}}).

The time periods within which high-resolution remote-sensing observations were taken for the Slow Wind SOOP ran from 2022 March 3 at 06:30~UT until 2022 March 6 at 18:30~UT (RSW1, 3.5 days in duration) and from 2022 March 17 06:00~UT to 2022 March 22 00:00~UT (RSW2, 5 days) when Solar Orbiter was at a distance from the Sun of 0.55--0.51 and 0.38--0.34~au, respectively. Solar Orbiter was close to the Sun-Earth line for the first instance and had a separation angle of 27-48$^\circ$ from Earth for the second. Observations were taken using all ten instruments on board. The various studies and modes of the instruments are shown in Figure~\ref{fig:fig1}~b and are described below. These can mainly be split into continuous, synoptic and high-resolution observations. The SPectral Imaging of the Coronal Environment (SPICE; \citealt{spice2020,fludra2021}) instrument is central to achieving the SOOP science goal and so this instrument and its observing modes are described first, followed by the rest of the remote sensing, and finally the in situ instruments.

SPICE is an EUV imaging spectrometer that operates in two wavebands: 704--790~\AA~(short wavelengths), and 973--1049~\AA~(long wavelengths), providing coverage from the upper chromosphere to the low corona (20\,000~K to 2~MK) and two lines that cover the flaring corona (up to 10~MK). The SPICE instrument provides intensities and line widths. SPICE also allows for the investigation of plasma composition through the derivation of elemental abundances. It produces rasters, created by scanning the solar disk from west to east (right to left), acquiring either a full spectrum or a limited wavelength range over the course of a few hours. For the slow wind SOOP, SPICE operated a combination of the following three studies each day:

\begin{itemize}
    \item {\bf Spectral Atlas.} FOV 2'x11', 4" steps, 60~s exposure, 53~min duration, 1 per day, full spectrum,
    \item {\bf Composition Map.} FOV 10.6'x11', 4" steps, 60~s exposure, 2.7 hr duration, which includes the following lines: \ion{O}{3} 703~\AA~multiplet, \ion{Mg}{9} 706~\AA, \ion{S}{4} 750~\AA, \ion{N}{4} 765~\AA, \ion{Ne}{8} 770~\AA, \ion{Mg}{8} 772~\AA, \ion{S}{5} 78.6~\AA, \ion{O}{4} 787~\AA, \ion{C}{3} 977~\AA, \ion{O}{6} 1032~\AA,
    \item {\bf Slow Dynamics.} Synoptic study, FOV 12.8'x11', 4" steps, 30~s exposure, 1.6~hr duration for each repetition (5-6 times), which includes the following lines: 
    \ion{O}{3} 70.3~\AA~multiplet, \ion{Mg}{9} 706~\AA, \ion{O}{2} 718~\AA, \ion{O}{5} 760~\AA~multiplet, \ion{N}{4} 765~\AA, \ion{Ne}{8} 770~\AA, \ion{S}{5} 786~\AA, \ion{O}{4} 787~\AA, \ion{C}{3} 977~\AA, \ion{H}{1} 1025~\AA, \ion{O}{6} 1032~\AA. 
\end{itemize}

The observation IDs for SPICE data taken during RSW1 and RSW2 are 100663695 to 100663711, and, 100663837 to 100663864, respectively.

The observations taken by SPICE were complemented by both synoptic and high-resolution images from the Extreme Ultraviolet Imager (EUI; \citealt{rochus2020}) and the Polarimetric and Helioseismic Imager (PHI; \citealt{solanki2020}). The EUI instrument has three different telescopes; one Full Sun Imager (FSI) providing images in two passbands (174~\AA~ and 304~\AA~), and two high-resolution imagers that capture the Sun in 1216~\AA~(\ion{H}{1}Lyman-$\alpha$) and 174~\AA. The EUI therefore covers from the upper chromosphere up to the low corona. Both HRIs take 2048~$\times$~2048~pixel images with a temporal resolution of 2-30~s. The imagers, that have a FOV of 17\arcmin $\times$ 17\arcmin corresponding to 0.28~R$_{\odot}^{2}$ when the spacecraft is at perihelion, capture the small-scale structure of the slow wind source regions. During the two slow wind SOOP windows, FSI synoptics were taken with a 900~s cadence and high-resolution images were taken with a cadence of 6~s over a period of 1~hr per day. The EUI observations taken for all SOOPs that operated during the first close perihelion passage are described in \citet{berghmans2023}.

The PHI instrument consists of two telescopes: one that images the full Sun, known as the Full Disc Telescope (FDT), which has a FOV of 2$^{\circ}$ $\times$ 2$^{\circ}$ and an angular resolution of 3.75'' per pixel, and the high resolution telescope (HRT) with a FOV of 0.28$^{\circ}$ $\times$ 0.28$^{\circ}$ with an angular resolution of 0.5'' per pixel \citep{gandorfer2018}. PHI scans the Fe I 6173~\AA~spectral line to produce filtergrams at 6 different wavelength positions in all 4 Stokes parameters (i.e. full polarimetry). Raw data can be compressed and downlinked directly or the inversion can be carried out on board. PHI provides the continuum intensity, line-of-sight (LOS) velocity, LOS magnetic field, and the vector magnetic field. PHI observations can be used to understand the evolution of the source region's magnetic field, and, along with EUI, to determine whether interchange reconnection is taking place at the CH or AR boundary. Extrapolations of the photospheric field can also be constructed that provide the coronal magnetic field structure across the boundary of the source region. The FDT only operated during the second observing window, providing full disk images with a 360~min cadence. Conversely, the HRT operated with a 3 and 5~min cadence for 1~hr per day, during the first and second remote sensing windows, respectively, where these 1~hr high-resolution observation time windows coincide with the observations taken by the EUI/HRI telescopes. The aberrations introduced by the non-radial temperature gradient in the entrance window have been removed as described by Kahil et al. (2023, submitted to A\&A).

Metis \citep{antonucci2020} is a solar coronagraph that simultaneously images the outer solar corona in visible and UV wavelengths. As we observe on-disk targets and Metis is used for off-limb observations of the solar corona, we did not require Metis for these particular runs of the SOOP. Therefore, Metis took generic observations during the first window of the SOOP, but was closed during the second window. Finally, the Spectrometer Telescope for Imaging X-rays (STIX; \citealt{krucker2020}), that observes solar flare X-ray emission, and the Solar Orbiter Heliospheric Imager (SoloHI; \citealt{howard2020}), that provides white-light images of the outer solar corona, operated in normal modes.

%the in situ instruments EPD/MAG/RPW/SWA.
In coordination with the high-resolution remote sensing observations of the slow solar wind source regions, measurements of the solar wind plasma were made by the in situ payload comprising four instruments. Measurements of electrons, protons, alpha particles and heavy ions are made using the Solar Wind Analyser (SWA; \citealt{owen2020}). SWA consists of a suite of three sensors: the Electron Analyser System (EAS), the Proton-Alpha Sensor (PAS) and the Heavy Ion Sensor (HIS) which provide high-resolution, full 3D velocity distribution functions and bulk plasma parameters of electrons, major and minor ion species along with elemental abundances and charge states of heavy ions. PAS and EAS ran continuously in normal mode with short periods of burst mode having a cadence of 4~s and $<$~1~s, respectively. HIS ran in normal mode with a cadence of 5~min with burst periods with a cadence of $<$~30~s. The Solar Orbiter Magnetometer (MAG; \citealt{horbury2020}) ran in burst mode during the SOOP providing interplanetary field measurements with 64 vectors~s$^{-1}$, which are made available in radial-tangential-normal (RTN) and spacecraft (SRF) coordinates. The magnetic field, particularly the radial component, is essential in tracing the connectivity of the spacecraft back to the source region on the Sun. The Energetic Particle Detector (EPD; \citealt{Pacheco2020}) consists of a suite of sensors including: the Supra Thermal Electron and Proton (STEP) sensor, the Electron-Proton Telescope (EPT), the Suprathermal Ion Spectrograph (SIS), and the High-Energy Telescope (HET) in order to record the parameters of different populations of energetic particles. A summary of the measurement capabilities of EPD can be found in Table~2 of \citet{Pacheco2020}. EPD operated continuously in close burst and normal modes as Solar Orbiter was within 0.7~au, with 15~min per day allocated to burst mode. Finally, the Radio and Plasma Waves (RPW; \citealt{maksimovic2020}) instrument measures waves along with electric and magnetic fields, observing radio emissions up to 16~MHz. RPW operated in normal high mode with 720~min a day allocated to burst mode measurements.

\section{Magnetic Connectivity}

\begin{figure*}[ht!]
\plotone{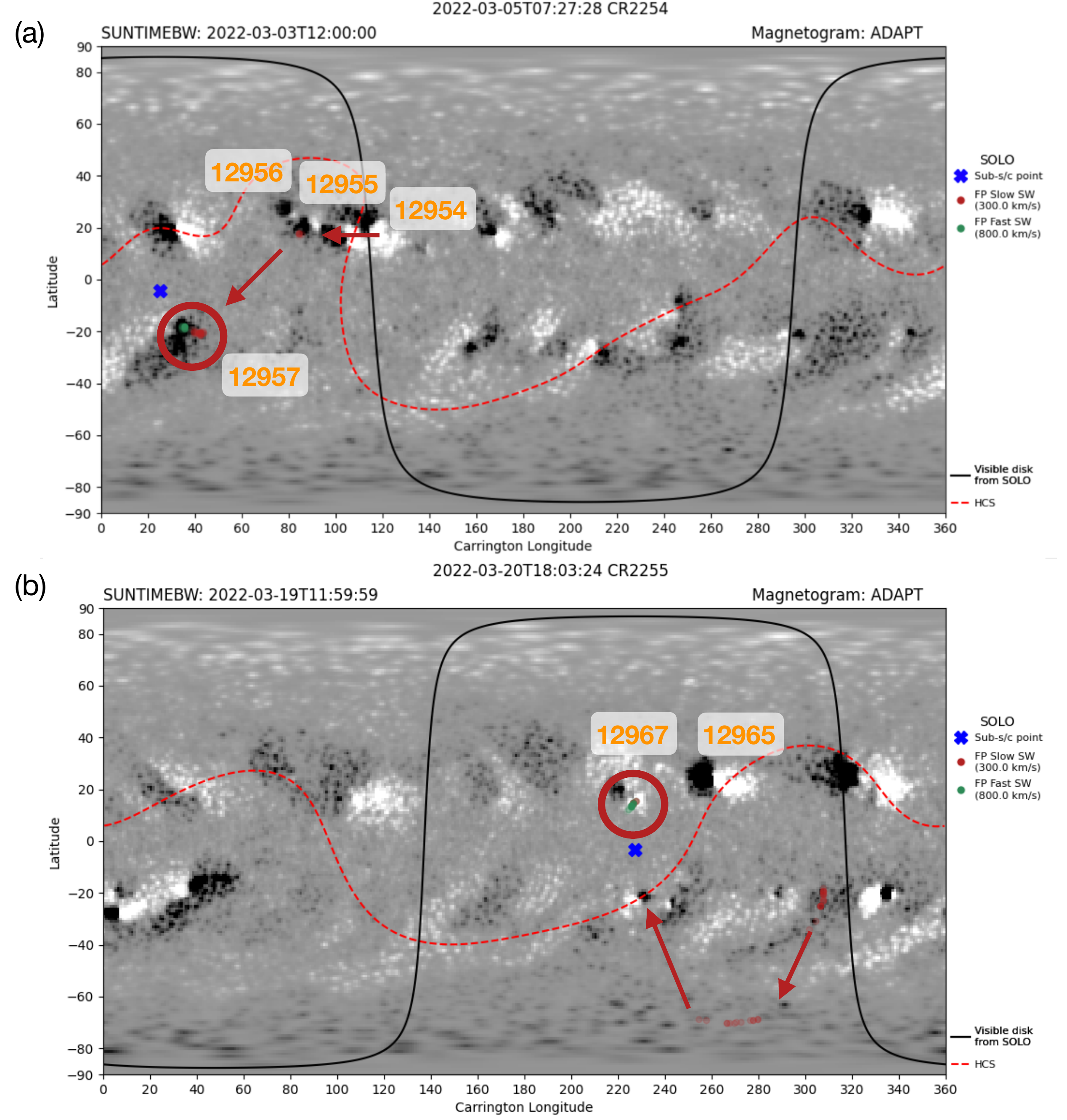}
\caption{Panel~a shows the predicted connectivity of Solar Orbiter traced back to the solar surface on an ADAPT magnetic field map at a Suntime (plasma release time) of 2022 March 3 at 12:00:00~UT, 6 hours after the start of the first remote sensing window, with a predicted solar wind release time of 2022 March 5 at 07:27:28~UT. The positions of the green and red connectivity points are calculated by assuming fast and slow solar wind speeds of 800 and 300~km~s$^{-1}$, respectively. The red arrows and circle show the evolution in the magnetic connectivity that occurs prior to the start of the remote sensing window that begins on 2022 March 3 at 06:00~UT. White (black) represents positive (negative) photospheric magnetic field polarities and the sub-spacecraft point is indicated by the blue cross. The black line encompasses the region of the solar disk that is visible from Solar Orbiter and the red dotted line shows the location of the HCS determined from the coronal field model. Panel~b shows the predicted connectivity of Solar Orbiter during RSW2. The labels are the same as panel~a. \label{fig:fig2}}
\end{figure*}

Tracing solar wind plasma measured in situ by Solar Orbiter back to the Sun and estimating the location of its source region requires modelling the magnetic connectivity of the spacecraft, which can be approximated using ballistic backmapping. In the first instance we assume the magnetic field structure to be constant over the time of flight of particles from the solar wind source to the satellite. In this case, particle trajectories are field-aligned and this will allow us to identify, in broad terms, areas in which the solar wind arriving at the spacecraft originated.  However, in general, field lines will move and the topology will evolve during the particle transit times. This means that although particles move along field lines, their trajectories are not field-aligned and will give rise to dispersive features in both energy and pitch angle. If there is a single reconnection location, these dispersions could be exploited to give much more detailed information from the particle measurements. An example of the application of such ideas has been in the cusp regions of Earth's magnetosphere, which are the outflow regions from the dayside magnetopause reconnection site: much has been deduced from these dispersions from the length and field strength distribution along the magnetic field lines (i.e., knowledge of the magnetic connectivity). For example, from the cusp ions of longest flight time, the conditions at the reconnection site and the reconnection rate have been derived \citep{lockwood1995, lockwood1995etal, lockwood1997, lockwood1998}. Application in the solar corona is more complicated due to the more complex field topology and because field lines may undergo a number of reconnection events, rather than at a single reconnection site giving rise to a distinct and identifiable particle population. Nevertheless, initial studies have already made deductions using particle dispersions of suprathermal electrons  in interplanetary space \citep{larson1997,owens2009}.

To model the connectivity of Solar Orbiter, the MADAWG connectivity tool\footnote{\url{http://connect-tool.irap.omp.eu/}} \citep{rouillard2020} is used. The connectivity tool combines different models of the coronal and interplanetary magnetic field to either forecast the magnetic connectivity of the spacecraft in advance, or to aid with post-observation analysis. Currently, by default, the tool reconstructs the coronal magnetic field using the potential field source surface (PFSS) model. The corona is then extended into interplanetary space by assuming a Parker spiral. In this case, the connectivity tool is used for both forecasting the magnetic connectivity of Solar Orbiter to determine the target pointing for the slow wind SOOP and also to aid with the connectivity analysis post-observation.

Immediately prior to the SOOP observing window, and throughout it, the SOOP coordinators used the tool to predict in advance the source region of the solar wind that will be detected at Solar Orbiter a few days later (i.e. accounting for Sun to spacecraft, Solar Wind lag), the coronal magnetic field is constructed using time-evolved synchronic ADAPT\footnote{\url{https://www.nso.edu/data/nisp-data/adapt-maps/}} \citep{arge2010,hickmann2015} magnetic field maps as the boundary condition. The tool then assumes two extreme values for the solar wind speed: 300~km~s$^{-1}$ for slow and 800~km~s$^{-1}$ for fast wind. These two speeds are used to construct the shape of the corresponding Parker spirals and the tool then computes the magnetic connectivity points at the source surface, and then at the solar surface, by tracing the coronal field. The tool provides a distribution of connectivity points, rather than a single point, to take into account uncertainties in the magnetic field tracing.

When applying the tool to aid with post-observation analysis to trace solar wind plasma, measured in situ at the spacecraft, back to its source on the Sun (spacecraft to Sun, SW lag) the propagation time needs to be calculated using observed speeds. Once the tool has calculated the propagation time of the solar wind parcel, the appropriate magnetogram that should be used as the boundary condition to reconstruct the coronal field is determined. The propagation time also defines the azimuthal drift of the connection points to the Sun, that keeps rotated while the solar wind plasma propagates. The Parker spiral is constructed using the solar wind speed measured by Solar Orbiter that has been retrieved either from low latency or fully calibrated science data from SWA/PAS. If the measured solar wind speed is unavailable, then the two extreme values (300, 800~km~s$^{-1}$) are used and the connectivity points are determined as described previously.

\subsection{Connectivity Prediction for Remote-sensing Window 1: 2022 March 3 to 6}

\begin{figure*}[ht!]
\plotone{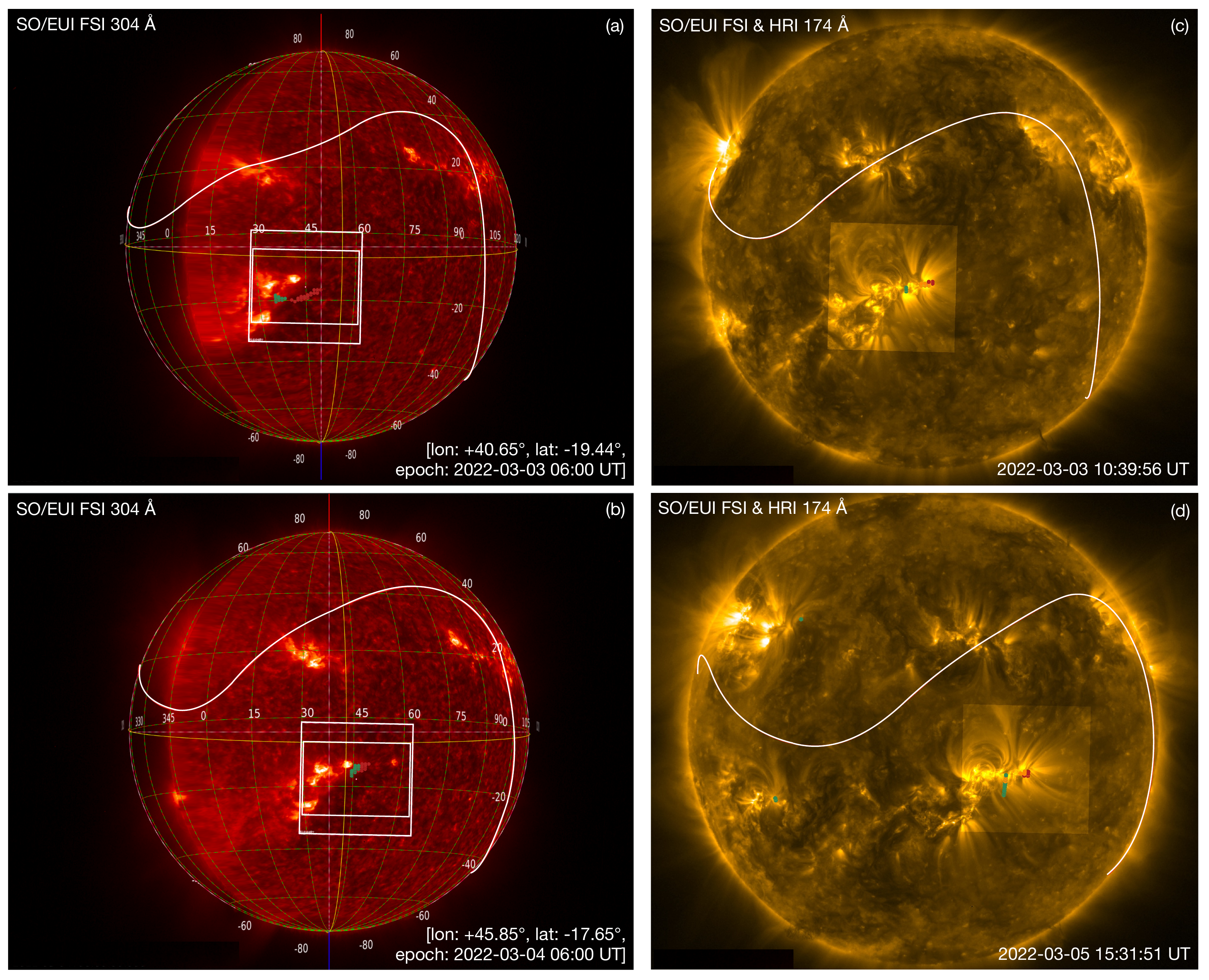}
\caption{The target selection for the high-resolution remote sensing instruments on board Solar Orbiter on 2022 March 3 and 5. The left panels show the low latency EUI/FSI 304~\AA~images using JHelioviewer \citep{muller2017} from 2022 February 27 at 08:50~UT and 2022 March 1 at 03:50~UT projected to 2022 March 3 at 06:00~UT and 2022 March 4 at 06:00~UT, respectively. The white boxes show the approximate size and location of the FOVs of EUI/HRI and PHI/HRT (large), and SPICE (small). The red (green) circles represent the distribution of connectivity points predicted for slow (fast) wind from the connectivity tool and the white line shows the position of the HCS. The grid labels are shown in Carrington coordinates. The text in the bottom right gives the feature triplet provided to Solar Orbiter, which includes Carrington latitude and longitude, and the epoch of the coordinates. The magnetic differential rotation model was applied to track the feature. The right panels show the resulting EUI/FSI and HRI 174~\AA~observations on 2022 March 3 and 5. The distribution of the post-observation connectivity points are also shown. \label{fig:fig3}} 
\end{figure*}

To predict the source region that Solar Orbiter would be magnetically connected to during RSW1, the connectivity was analysed both on long and short time scales. Firstly, we analysed the connectivity of Solar Orbiter (SW lag, from spacecraft to Sun) during the previous solar rotation (roughly between 2022 February 4 and 7) as both AR and CH solar wind source regions can be present for multiple rotations. The connectivity tool showed that the spacecraft was likely connected to the negative polarity of AR 12939 in the southern hemisphere. The solar wind speed measured by SWA/PAS on 2022 February 8 at 06:00~UT was $\sim$350~km~s$^{-1}$ but speeds of up to $\sim$500~km~s$^{-1}$ were measured a few days prior. The radial magnetic field measured by MAG is directed inwards, which agrees with the negative polarity source region given by the connectivity tool.

The connectivity tool was also used to forecast the connectivity of Solar Orbiter (Sun to spacecraft, SW lag) in the days leading up to the observation windows. The target of the high-resolution remote sensing instruments on board Solar Orbiter is decided roughly three days in advance of the observations taking place during very short term planning (VSTP) or pointing decision meetings. For the first window there were three meetings, each a day apart, held on February 28 until March 2. Therefore, once the initial pointing had been decided there were two additional opportunities to update the target (p-VSTP update). The connectivity of Solar Orbiter was monitored before and during the pointing decision meetings. The connectivity tool predicted that the spacecraft would be magnetically connected to the positive polarity of the AR in the northern hemisphere before the start of the window (see Figure~\ref{fig:fig2}~a). Then there would be a transition in the connectivity across the heliospheric current sheet (HCS) towards the negative polarity of the same AR and then to the east to the smaller AR. Finally, it would move across to the decayed negative polarity of AR 12939, which was present during the previous rotation and was now AR 12957 during the rotation of interest, and remain stable for several days. The evolution of the connectivity is indicated by the red arrows in Figure~\ref{fig:fig2}.

\subsection{Connectivity Prediction for Remote Sensing Window 2: 2022 March 17 to 22}

Predicting the magnetic connectivity of Solar Orbiter in advance of the second remote sensing window (RSW2) of the Slow Wind SOOP was more difficult compared to the first window. As RSW2 approached it became clear that the connectivity of Solar Orbiter would be different to that of the previous rotation. Also, it was apparent that the connectivity would be more complex and variable during this window compared to RSW1. The evolution of the predicted magnetic connectivity of Solar Orbiter for RSW2 can be seen in Figure~\ref{fig:fig2}~b. Prior to the pointing decision meetings and RSW2, the initial target suggested by the connectivity tool, was the negative polarity of AR 12965 located in the northern hemisphere (adjacent to the red circle in Figure~\ref{fig:fig2}~b), with the connectivity remaining relatively stable over the 5-day window. However, on the day prior to the pointing decision meetings for RSW2, which were held between 2022 March 17 and 18, the predicted connectivity changed dramatically. The rapid change in the connectivity prediction for Solar Orbiter was due to the increased complexity of the Sun's magnetic field compared to RSW1. Before RSW2 there were multiple ARs either emerging or rotating onto the visible disk. Also, the low latency MAG data showed the arrival of a CME during March 11-12, which made it impossible to determine the background magnetic field direction during the initial pointing decision meeting.

Initially, at the start of the observing window (Suntime 2022 March 17 at 05:59~UT), the connectivity tool forecasts that Solar Orbiter will be primarily connected to the decayed negative polarity close to the west limb as observed by the spacecraft. Shortly afterwards, on March 19, the location of the connectivity points transitioned to the southern polar coronal hole, also consisting of dominantly negative polarity magnetic field. Finally, the connectivity points cross the HCS and, on March 20, Solar Orbiter is connected to the positive polarity of small, decaying AR 12967. The connectivity points were predicted to remain in this location for the last two days of RSW2. 

However, on the basis of the connectivity location predicted immediately prior to RSW2, the initial target was chosen to be the boundary of the southern polar coronal hole during the initial two pointing decision meetings (see Section~\ref{sec:targetRSW2}), the predicted connectivity and shape of the HCS changed significantly. It was therefore challenging to select the target for the second half of RSW2. The predicted connectivity and shape of the HCS changed significantly due to substantial flux emergence occurring in the large AR (12965, see Figure~\ref{fig:fig2}~b) in the northern hemisphere. This substantial emergence was apparent in the ADAPT magnetic field maps and affected the coronal magnetic field model. This led to a watch-and-wait approach in terms of target selection until the large-scale emergence of the AR no longer influenced the connectivity.

\section{Target Selection for RSWs 1 \& 2} \label{sec:target}

\subsection{Target Pointing RSW1: 2022 March 3 to 6}

\begin{figure*}[ht!]
\plotone{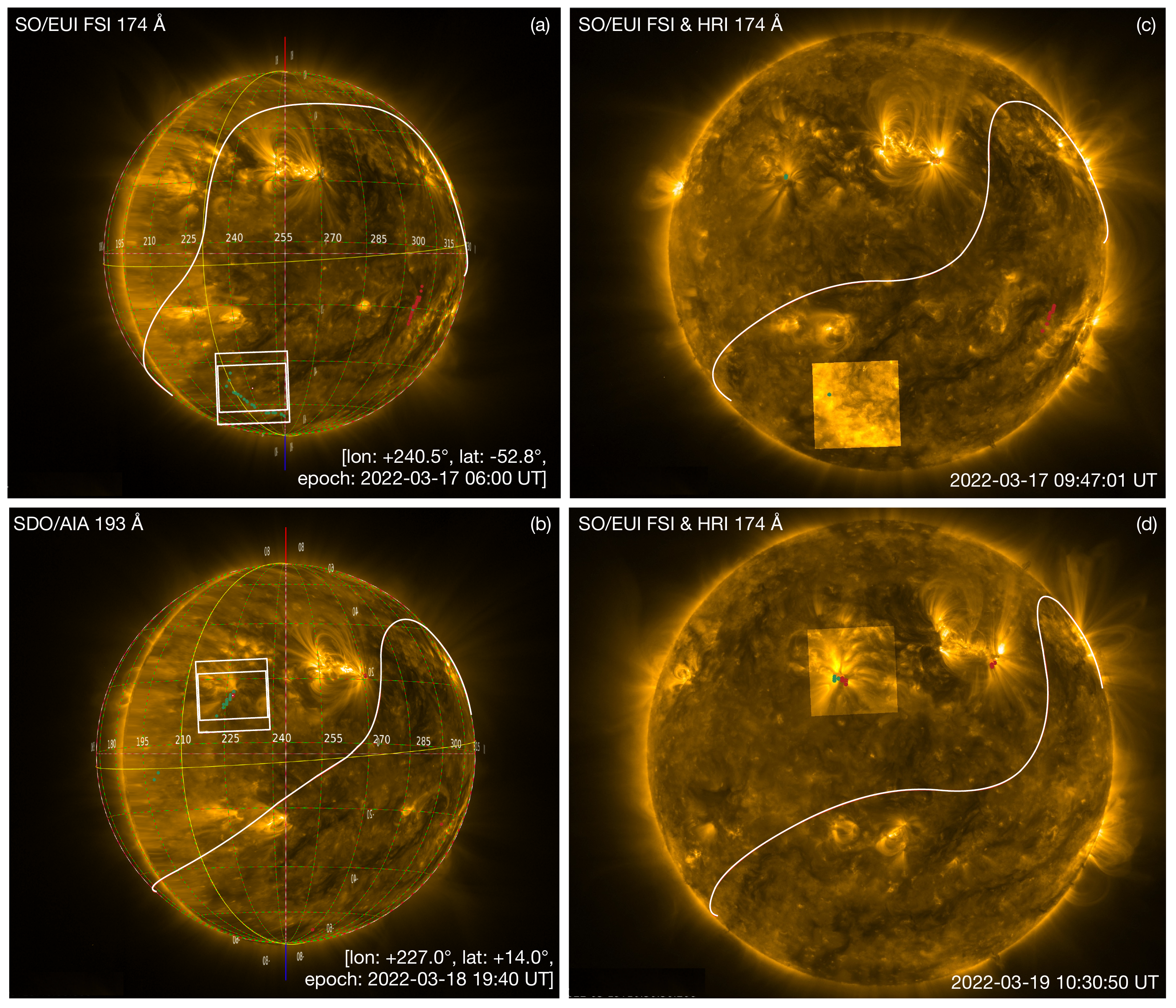}
\caption{The target selection for the high-resolution observations taking during RSW2 on 2022 March 17 and 18. The left panels show the low latency EUI/FSI 174~\AA~images from 2022 March 13 at 10:30~UT and March 15 at 08:30~UT, respectively. The right panels show the EUI/FSI and HRI 174~\AA~observations taken on 2022 March 17 and 19. The labels are the same as described in Figure~\ref{fig:fig3}.  } \label{fig:fig4}
\end{figure*}

The left panels of Figure~\ref{fig:fig3} show the FOVs resulting from two target pointing decisions made before RSW1 of the Slow Wind SOOP on EUI/FSI low latency 304~\AA\ image. The pointing decision was initially applied to all p-VSTP updates during RSW1 (Figure~\ref{fig:fig3}~a). The pointing was adjusted on March 1, which came into effect on March 3 at 18:00~UT, in order to follow the predicted location of the connectivity points (Figure~\ref{fig:fig3}~b). This second pointing decision was maintained during the rest of the first window. The target for RSW1 was the boundary of AR 12957, which was adjacent to an equatorial coronal hole, as the connectivity tool had predicted that Solar Orbiter would be connected to this region.

Figure~\ref{fig:fig3}~c,d show the full disk 174~\AA~images taken by EUI/FSI along with the high-resolution EUI/HRI 174~\AA~and post-observation connectivity points superimposed. High-resolution data were taken by EUI and PHI for roughly one hour periods during 2022 March 3 from 09:40 to 10:40~UT, 2022 March 4 from 10:45 to 11:45~UT, 2022 March 5 from 15:05 to 16:05~UT, while SPICE operated continuously. The points from the connectivity tool, along with the remote sensing data, suggest that images of the slow solar wind source region were successfully captured during the 1-hr high-resolution observation windows on 2022 March 3 and 5.

\begin{figure*}[ht!]
\plotone{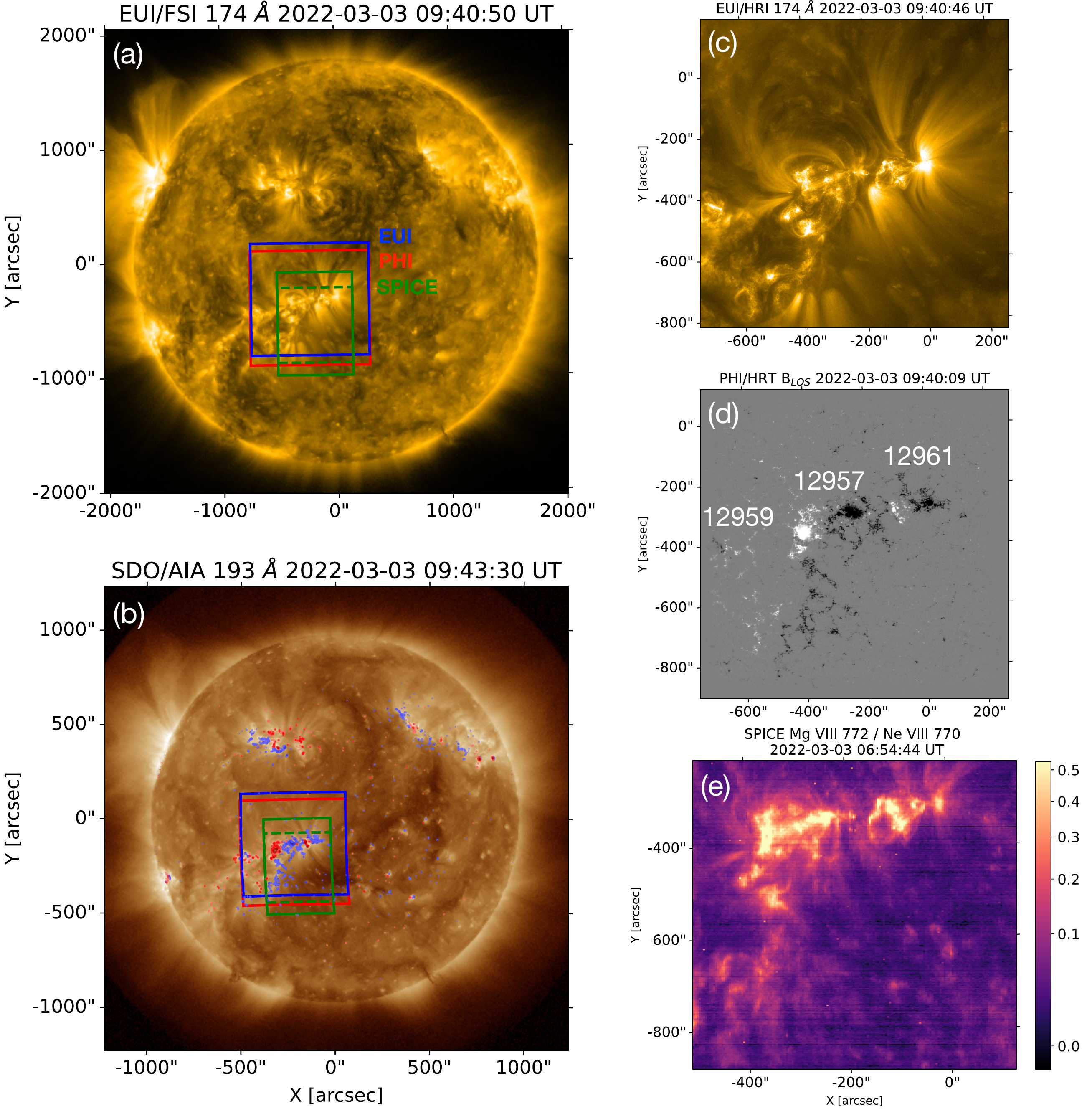}
\caption{Remote sensing data taken at the beginning of RSW1 of the Slow Wind SOOP. Panel~a shows an EUI/FSI 174~\AA~image taken at the start of the first 1-hr high-resolution observation window. The FOVs of EUI/HRI, PHI/HRT, and SPICE are given by the blue, red, green boxes, respectively. The green dashed box shows the FOV of the SPICE data in panel~e cropped to remove the bright dumbbell at the top of the raster. Panel~b shows the FOVs on the corresponding 193~\AA~image taken by the Atmospheric Imaging Assembly \citep{lemen2012} onboard SDO \citep{pesnell2012} with photospheric magnetic polarities overlaid from the Helioseismic and Magnetic Imager \citep{scherrer2012} where red (blue) represents positive (negative) field with a saturation of $\pm$300~G. Panels~c--e show the EUI/HRI 174~\AA , the PHI/HRT line-of-sight magnetic field and SPICE \ion{Mg}{8}/\ion{Ne}{8} spectral window at wavelength position 23. In panel~d, white (black) represents positive (negative) field with a saturation level of $\pm$50~G and ARs 12957, 12959, and 12961 are labelled in white. \label{fig:fig5}} 
\end{figure*}

\subsection{Target Pointing RSW2: 2022 March 17 to 22} \label{sec:targetRSW2}

The left panels of Figure~\ref{fig:fig4} show the FOVs resulting from the targets chosen from two pointing decisions made in advance of RSW2. For RSW2, two different targets were selected due to the change in the predicted connectivity of Solar Orbiter during this window. The first target of RSW2 was selected to be the boundary of the southern polar coronal hole (Figure~\ref{fig:fig4}~a). Solar Orbiter observed the coronal hole boundary between 2022 March 17 at 06:00~UT to 2022 March 18 at 19:40~UT. The second target of RSW2 was chosen to be the positive polarity of AR 12967, located in the northern hemisphere (Figure~\ref{fig:fig4}~b), which was observed from 2022 March 18 at 19:40~UT until 2022 March 22 at 00:00~UT.

During the first pointing decision meeting on 2022 March 13, the first target of RSW2 was chosen as the boundary of the southern polar coronal hole, based on the predicted location of the fast wind connectivity points (Figure~\ref{fig:fig4}~c). This was due to the location of the predicted slow wind connectivity points being very close to the west limb in decayed negative field, adjacent to the predominantly closed field of a filament channel. After the first meeting, there were multiple possible targets as the connectivity changed drastically due to the large emerging AR in the northern hemisphere (Figure~\ref{fig:fig2}~b). Therefore, it was decided that the target pointing would remain the same until there was a clear target. By the third meeting, the connectivity had changed again and it was predicted that Solar Orbiter would be connected to the large positive polarity of AR 12965, switching to the positive polarity of AR 12967, the latter of which was chosen as the target for the rest of the window (Figure~\ref{fig:fig4}~d).  

\begin{figure*}[ht!]
\plotone{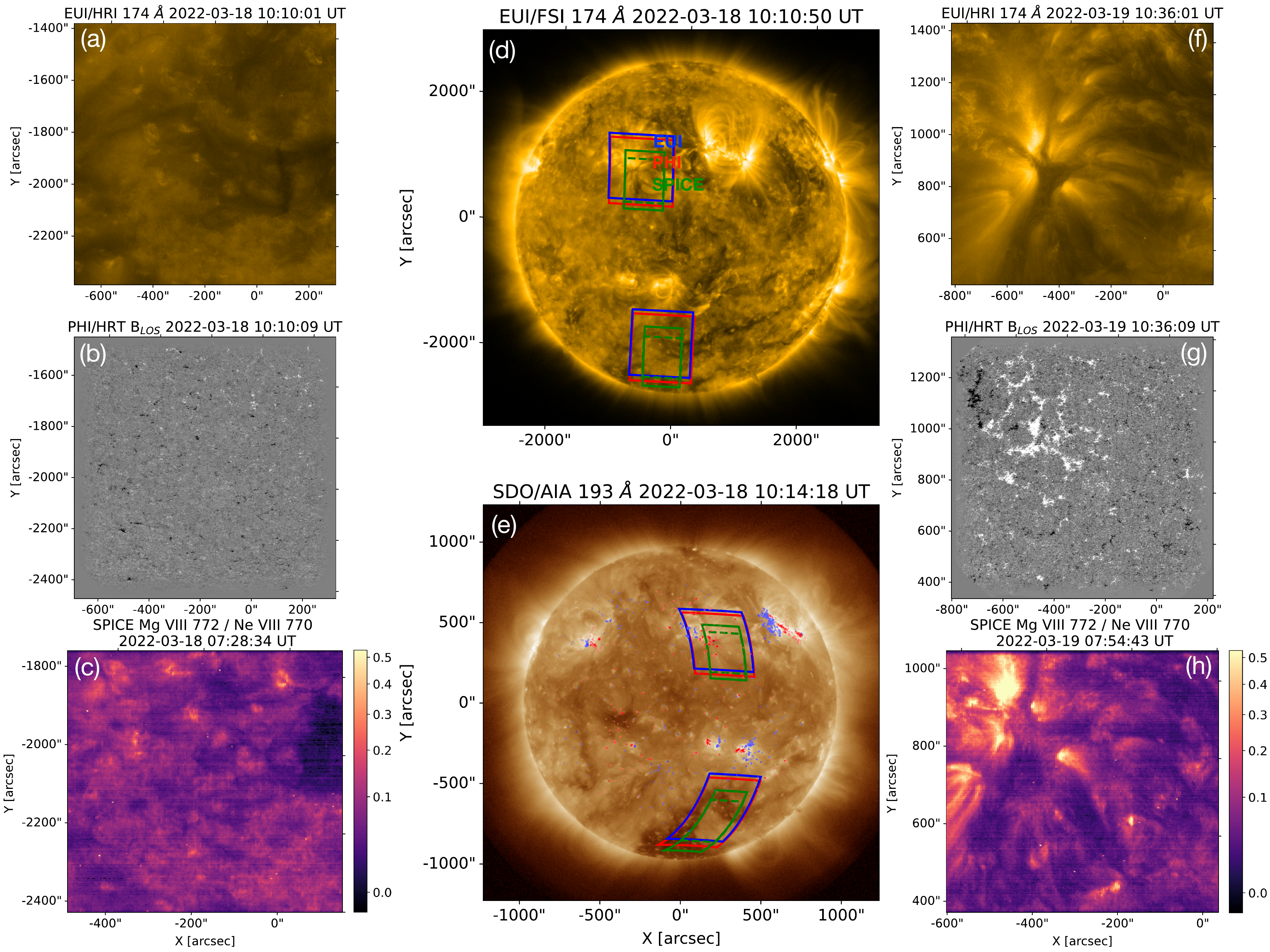}
\caption{High-resolution data taken during RSW2 of the Slow Wind SOOP on 2022 March 18 (panels~a--c) and 2022 March 19 (panels~f--h). The FOVs of EUI/HRI, PHI/HRT and SPICE are shown in blue, green, and red on EUI/FSI 174~\AA~and SDO/AIA 193~\AA~data, similar to Figure~\ref{fig:fig5}. \label{fig:fig6}} 
\end{figure*}

The right panels of Figure~\ref{fig:fig4} show the EUI/FSI and HRI observations taken at the start of two of the high-resolution remote sensing windows on 2022 March 17 and 19. There were only a few connectivity points captured within the EUI/HRI FOV. In Figure~\ref{fig:fig4}~d, the post-observation connectivity points are located at the footpoints of fan loops associated with AR 12967.

\section{High-resolution Observations}

\subsection{Remote Sensing Window 1}

%AR complex- source region
As previously stated, the target during RSW1 of the Slow Wind SOOP was the boundary of AR 12957, which evolves to become part of an AR complex composed of ARs 12957, 12959, and 12961. Figure~\ref{fig:fig5} shows the high-resolution EUI/HRI, PHI/HRT and SPICE data taken of target AR 12957 on 2022 March 3. AR 12957 was first visible on the east limb on 2022 February 24, as observed by STEREO-A, followed by Solar Orbiter and the Solar Dynamics Observatory (SDO) on 2022 February 25. As AR 12957 rotated onto the solar disk an equatorial coronal hole, predominantly consisting of small-scale negative polarity field, formed between the boundary of the AR, stretching across central meridian, towards the boundary of AR 12956 in the northern hemisphere (Figure~\ref{fig:fig5}~b). Late on March 3 the equatorial hole expands before shrinking in size on March 5. AR 12957 continuously emerges during its disk passage and on March 4 it evolves to become part of an AR complex including ARs 12959 and 12961 (Figure~\ref{fig:fig5}~d). 

This AR complex produced some eruptive activity, mainly in the time period leading up to RSW1. On 2022 March 2 at around ~08:00~UT there was a C3.1 class flare that was recorded both by STIX and GOES, and a CME observed by STEREO-A/COR2 and LASCO/C2. The eruption originated from the southern part of AR 12959 from an area of decayed field, away from the region Solar Orbiter was connected to. There was another eruption from this complex originating from AR 12957 on 2022 March 2 at 09:30~UT, with another C3.1 class flare. Also, a filament eruption occurred on 2022 March 5 at 15:15~UT originating from AR 12959 associated with a C1.1 flare. 

%2022 march 03
During the 1-hr high-resolution observations on 2022 March 3, coronal fan loops are visible extending from the edge of the negative polarities of ARs 12957 and 12961 (see Figure~\ref{fig:fig5}~c,d). EUV brightenings and flows are observed in AR core loops, while jets are visible at the boundary of the negative polarity of AR 12597 and the edge of the positive polarity of AR 12961, at around 10:06~UT and 10:19~UT, respectively. The jets appear to be helical in nature, with material propagating along closed rather than open loops.

\subsection{Remote Sensing Window 2}

Figure~\ref{fig:fig6} shows high-resolution observations of the two different targets taken during RSW2 on 2022 March 18 and 19. Figure~\ref{fig:fig6}~a--c show the EUI/HRI, PHI/HRT and SPICE observations taken of the first target consisting of the polar coronal hole boundary and a filament channel in the southern hemisphere. During the 1-hr high-resolution observation window on 2022 March 18 the filament erupts and the associated coronal dimming expands and merges with the southern polar coronal hole. The filament eruptions begins around 02:30~UT on March 18. The coronal dimming associated with the filament eruption expands towards and eventually starts to merge with the southern polar coronal hole at approximately 08:30~UT and continues until the target is updated on March 18 at 19:40~UT. The merging of the coronal dimming, associated with the filament eruption, with the southern polar coronal hole are investigated in \citet{nawin2023}.

Figure~\ref{fig:fig6}~f--h show the high-resolution remote sensing observations of the second target of RSW2 on March 19. The target for the remainder of the window was the positive polarity of AR 12967. The AR was first visible on the solar disk as viewed by SDO on March 12 and the leading positive polarity spot had already decayed by the time of the Solar Orbiter observations on March 19. AR 12967 produced an eruption on March 20 at 07:10~UT and another smaller eruption on March 21 at $\sim$10:48~UT. The AR continues to decay as it rotates across the solar disk out of view. \citet{baker2023} investigates the evolution of a thin corridor of upflows adjacent to the AR using SDO and Hinode/EIS observations as a source region of S-web slow solar wind \citep{antiochos2011} The in situ observations from Solar Orbiter along with the magnetic connectivity tool are used to determine the connectivity evolution of Solar Orbiter across the corridor and AR.

\section{Coordinated Remote Sensing Observations}

\begin{figure*}[ht!]
\plotone{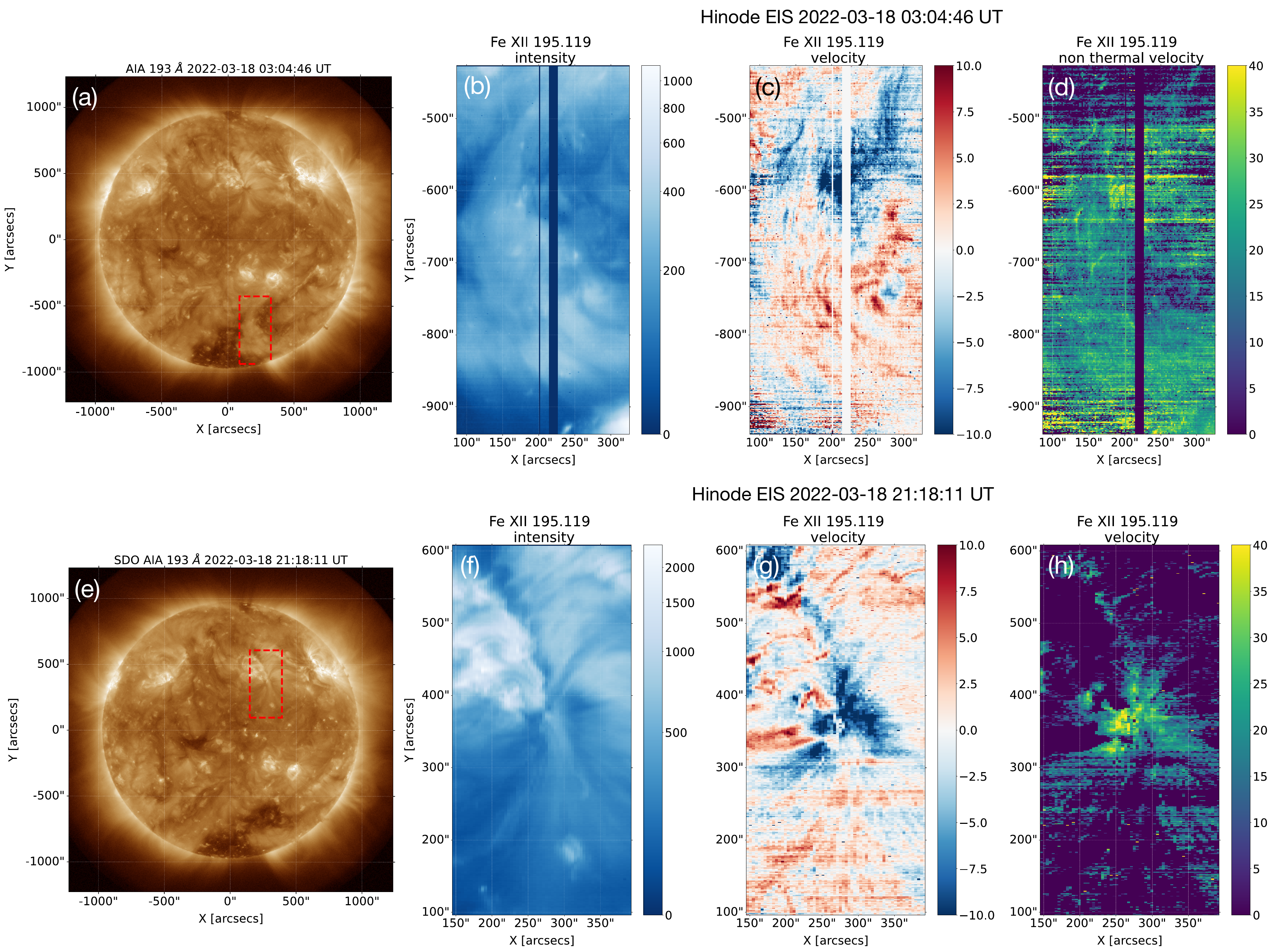}
\caption{Hinode/EIS observations taken during RSW2 of the Slow Wind SOOP. The panels from left to right show the SDO/AIA 193~\AA~images with the red dashed box showing the target and FOV for Hinode/EIS, Fe XII intensity, Doppler velocity and non-thermal velocity maps. In panel~h the non-thermal velocity maps only show pixels with negative Doppler velocity. \label{fig:fig7}} 
\end{figure*}

Coordinated observations were taken with Hinode and the Interface Region Imaging Spectrometer (IRIS; \citealt{depontieu2014,depontieu2021}) via IHOPs 433 and 434 during RSW1 and RSW2 of the Slow Wind SOOP. Coordinates and timings for the targets were provided for Hinode and IRIS using the Solar Orbiter planning website\footnote{\url{https://umbra.nascom.nasa.gov/solar\_orbiter/planning/}}.

\subsection{Hinode/EIS}
EIS is a normal incidence spectrometer that observes the 171--211\,\AA\, and 245--291\,\AA\, wavelength bands. Details of the instrument are given in \cite{Culhane2007}. The instrument has a spectral resolution of 23\,m\AA, and can observe with one of four (1$''$, 2$''$, 40$''$, 266$''$) slits that are interchangeable via a narrow-wide slit assembly. EIS has a moderate spatial resolution with 1$''$ spatial pixels, and takes exposures of tens of seconds, depending on the science goal. A very wide range of spectral lines are available in the observational wavebands, providing numerous diagnostics of Doppler and non-thermal velocities, temperatures, densities, elemental abundances, and even coronal magnetic field strengths. 

Hinode was off-line in safe-hold mode just prior to the RSW1. Although observations had resumed by the time of the SOOP, no data dedicated to HOP 434 were taken. Nevertheless, some observations of the target AR were obtained, and could be useful for complementary studies to the work presented here. 

Support for IHOPs 433/434 was much more extensive during RSW2. EIS ran a combination of the following four studies:
\begin{itemize}
    \item {\bf Atlas\_60.} Full CCD spectrum (171--211\,\AA\, and 245-291\,\AA), FOV 120$''$x160$''$, 2$''$ slit option, 60\,s exposures, $\sim$65\,min duration.
    \item {\bf DHB\_007\_v2.} Abundance map. FOV 248$''$x512$''$, 4$''$ steps with the 2$''$ slit option, 60\,s exposures, $\sim$68\,min duration, includes spectral lines from consecutive ionisation stages of Fe VIII--XVI and Ca XIV--XVII for deriving the coronal temperature and density, and the Si X 258.375\,\AA\, and S X 264.223\,\AA\, lines for measuring the Si/S elemental abundance ratio.
    \item {\bf HPW021VEL240x512v2\_b.} Spectroscopic diagnostic scan. FOV 240$''$x512$''$, 2$''$ steps with the 1$''$ slit option, 100\,s exposures, $\sim$3\,hr 34\,min duration, includes a wide range of spectral lines from Fe VIII--XXIV, hot Ca XIV--XVII lines, lower temperature O IV--O VI and Mg V--VII lines, abundance diagnostics from Si/S and Ca/Ar. 
    \item {\bf HPW024\_266fullccd25s.} Wide-slit full CCD movies. FOV 266$''$x256$''$, sit-and-stare with the 266$''$ slit option, 25\,s exposures with a 35\,s delay to make 1\, min cadence, duration depends on the number of repetitions. 
\end{itemize}
Typically the full CCD spectral atlas was run once per day, with diagnostic scans repeated throughout the rest of the available time-slots. A selection of the Hinode/EIS observations taken during RSW2 are shown in Figure~\ref{fig:fig7}. Noticeable features include strong upflows present in both Doppler maps (Figure~\ref{fig:fig7}~c and g). The region of strong upflows in Figure~\ref{fig:fig7}~c is the result of the expansion of a coronal dimming associated with an eruptive filament that merges with the southern polar coronal hole, which is the topic of \citet{nawin2023}. The second region of strong upflows in Figure~\ref{fig:fig7}~g is due to the existence of a narrow open-field corridor that is associated with the positive polarity of AR 12967 and extends towards the north that is magnetically connected to Solar Orbiter during the second half of RSW2 \citep{baker2023}.

\subsection{Hinode/XRT}
The X-ray Telescope (XRT) onboard Hinode \citep{golub2007} is a high-resolution, high cadence grazing incidence telescope. It observes the solar corona through nine X-ray filters on two filter wheels, which cover the wavelength range of 6--60~\AA~. This enables XRT to image plasma across a temperature range of 1 to $\sim$20~MK with a pixel size of 1\arcsec. XRT is therefore capable of imaging a broad range of plasma temperatures, and has a large dynamic range allowing a variety of coronal features to be detected. XRT takes full-Sun and partial-Sun field of view images. Coronal plasma temperatures and differential emission measures can be computed using filter ratio techniques. All data preparation and analysis techniques are detailed in the XRT User Guide\footnote{\url{https://xrt.cfa.harvard.edu/resources/documents/
XAG/XAG.pdf}}.

RSW1 was supported by XRT with data on March 3, 4, 5 and 6. Data taken during this time period include both partial frame images with AR 12957 in the field of view and full-sun images. Data taken with the following filters is available: Thin Al poly, Thin Al mesh, Medium Al, Thin Be, Medium Be, Thick Be. During RSW2, pointings are used for the partial frame images. First, the boundary of the southern polar coronal hole was the target at various times between 2022 March 17 at 06:00~UT to March 18 at 19:40~UT. The coronal dimming that merges with the coronal hole is particularly visible in the XRT Al poly filter \citep{nawin2023}. Second, between 2022 March 18 19:40~UT until the end of March 21 the positive polarity of AR 12967 in the northern hemisphere was the pointing selection. Full-sun images are also available during the RSW2. Data are collected using the Thin Al poly, Thin Al mesh and Thin Be filters.

\begin{figure*}[ht!]
\includegraphics[width=1\textwidth]{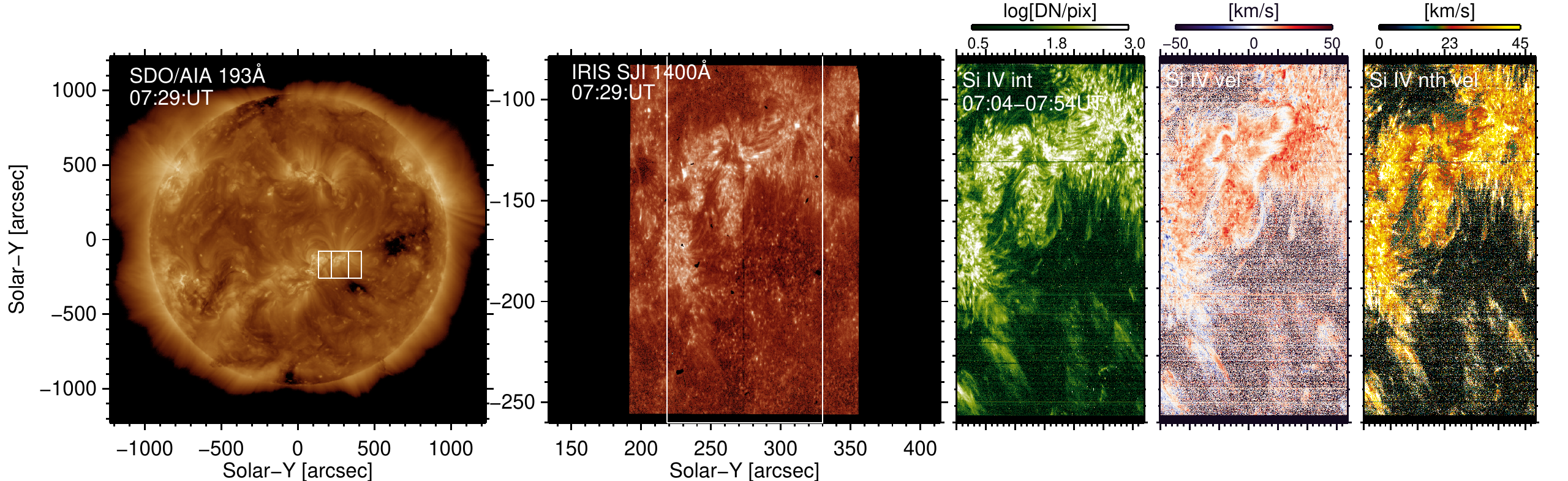}
\includegraphics[width=1\textwidth]{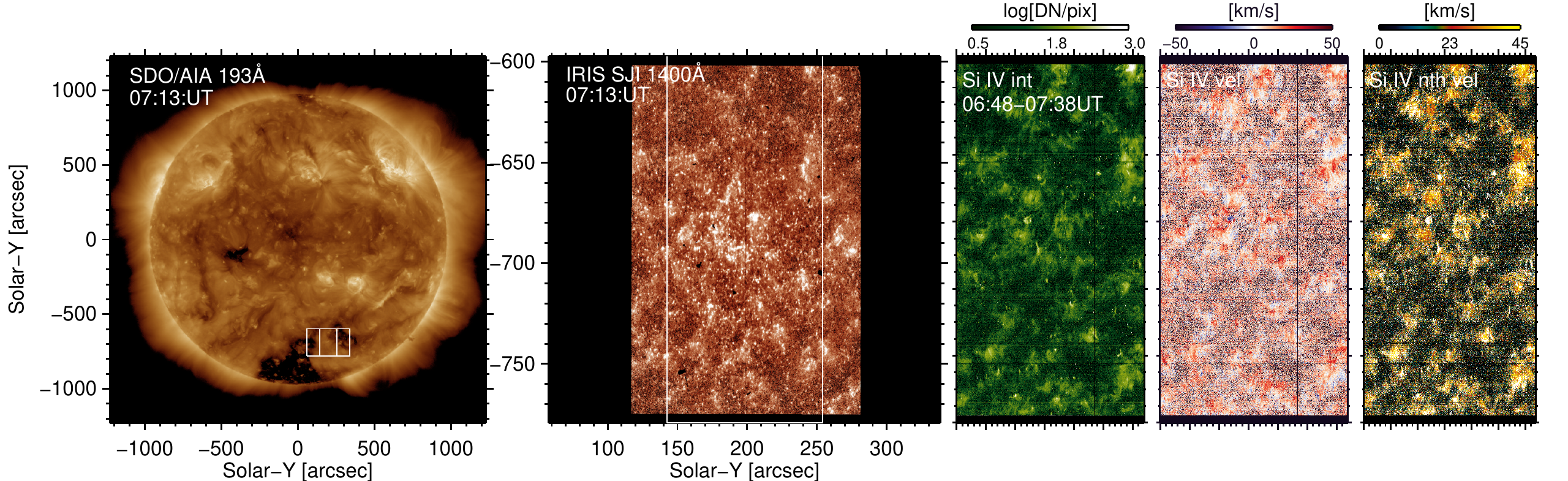}
\includegraphics[width=1\textwidth]{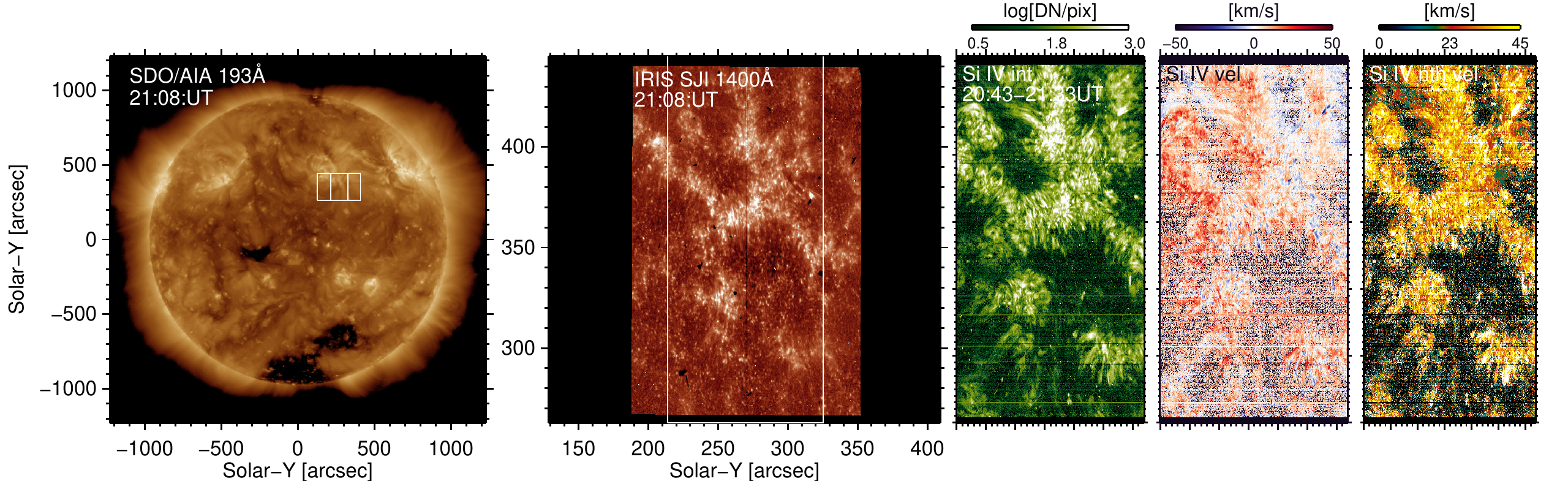}
\caption{Overview of IRIS observations during the RSW1 and RSW2 on 2022 March 3 and 18. The panels show (from left to right): a full-disk context AIA 193~\AA~image with the FOV of the IRIS SJI and spectrograph raster overlaid, a SJI in the 1400~\AA~filter with the spectrograph raster FOV overlaid and images of \ion{Si}{4} intensity, Doppler shift velocity and non-thermal velocity. } \label{fig:fig8}
\end{figure*}

\subsection{Hinode/SOT}
The Solar Optical Telescope (SOT; \citealt{tsuneta2008}) on board Hinode consists of a 50~cm diffraction-limited Gregorgian telescope known as the Optical Telescope Assembly, narrowband and broadband filtergraphs and the Stokes Spectro-Polarimeter (SP) which make up the Focal Plane Package. SOT/SP operates in a fixed band of wavelengths centered around the Fe I lines. In fast map mode, a 160'' scan takes SP 30~min with a polarimetric accuracy of 0.1\% with a resolution of 0.32''. SOT/SP took fast maps of different targets during RSW2 for both coordination with the Slow Wind SOOP and also for cross-calibration purposes with PHI. From March 18 fast maps with a FOV 32$''$ $\times$ 164$''$ and a duration of 30~min were taken near the southern polar coronal hole. On March 19 the target changed to AR 12965, followed by AR 12967 on March 20 until March 22.

\subsection{IRIS}
IRIS is a remote-sensing dual telescope and spectrograph acquiring images and spectra in the far (FUV) and near (NUV) ultraviolet  at very high spatial (0.33--0.4$''$),  temporal (down to $\sim$1~s) and spectral (2.7~km~s$^{-1}$ pixels) resolution. The IRIS' spectrograph observes continua and emission lines formed over a very broad range of temperatures, from ${\rm log} (T[K])=[3.7,7]$, including key spectral lines which provide unique diagnostics of plasma dynamics in the chromosphere (e.g. \ion{C}{2} and \ion{Mg}{2}) and transition region (e.g. \ion{Si}{4}). Simultaneously, the IRIS Slit Jaw Imager (SJI) provides high-resolution images in four different filters (\ion{C}{2} 1330~\AA, \ion{Si}{4} 1400~\AA, \ion{Mg}{2} k 2796~\AA~and \ion{Mg}{2} wing 2830~\AA).
Thanks to its unique combination of sub-arcsecond and high-cadence imaging spectroscopy, IRIS allows the trace of mass and energy transfer through the complex interface region between the photosphere and the corona, which is crucial to understand the origin of the solar wind \citep{depontieu2014,depontieu2021}.

IRIS supported the SOOP coordination during the period covering 2022 March 3--6 and March 17--22 leading up to Solar Orbiter's first perihelion passage, and ran a combination of the following observational programs:

\begin{itemize}
    \item \textbf{OBSID 3620259477}: very large dense (with two consecutive slits separated by 0.35'')  320-step rasters covering a field of view of 112''x175'' in about 50~min with an exposure time of 8~s. SJI images in the  1400~\AA~and 2796~\AA~filters were also taken with a cadence of 19~s. \textbf{OBSID 3600109477} with a spatial and spectral summing by 2 was also used as a low datarate alternative. 
    \item \textbf{OBSID 3600607418}: Very large dense 4-step raster with a FOV of 1''x175'', raster cadence of about 20~s and exposure time of 5~s. SJI images in the 1400~\AA~and 2796~\AA~were also acquired every 10~s. The observations were summed spatially by 2 and spectrally by 2 and 4 for the NUV and FUV channels respectively.
    \item \textbf{OBSID 3600609521, 3600609621}: both are very large sparse 4-step rasters covering a FOV of 3''x175'' in 37~s with an exposure time of 8~s. SJI images in the 1330~\AA~filter (OBSID 3600609521) or 1400~\AA (OBSID 3600609621) are taken every 9~s.
\item \textbf{OBSID 3600010076}: a large dense 320-raster over a FOV of 112''x120'' in about 90~min, with exposure time of 15~s. SJI images in the four filters were also taken with a cadence of 67~s.
\end{itemize} 

The majority of the observations were taken using the first two OBSIDs (and the low-telemetry alternative) from the list above. IRIS was following the same targets observed by Solar Orbiter, after compensating for the difference between the two instruments' position as compared to the Sun.

In Figure~\ref{fig:fig8} we show an overview of IRIS observations of three targets of interest (ARs in the top and bottom panels and a coronal hole boundary in the middle panels) during the RSW1 and RSW2. Each figure shows (from left to right): a full-Sun context image from the SDO/AIA 193~\AA~channel overlaid with the FOV of the IRIS SJI and raster, an IRIS SJI in the 1400~\AA~filter with the raster FOV overlaid, and finally images of \ion{Si}{4}~1394~\AA~ intensity, Doppler shift velocity and non-thermal velocity. The fit of the \ion{Si}{4} line was obtained by performing a single Gaussian function in each pixel using the \textit{iris$\_$auto$\_$fit.pro} IDL routine written by Peter Young\footnote{\url{http://www.pyoung.org/quick$\_$guides/iris$\_$auto$\_$fit.html}}. We chose to fit the strongest \ion{Si}{4}~1394~\AA~line in order to maximize the signal-to-noise ratio. This is particularly important for the coronal hole boundary observation (middle row of Figure~\ref{fig:fig8}), where the transition region lines are usually faint. To calculate the Doppler velocity, we used a reference wavelength of 1393.755~\AA, \citep[from CHIANTI][]{dere1997,delzanna2021} and the non-thermal velocity was calculated using the full-width at half maximum of the single Gaussian fit.

\begin{figure*}[ht!]
\plotone{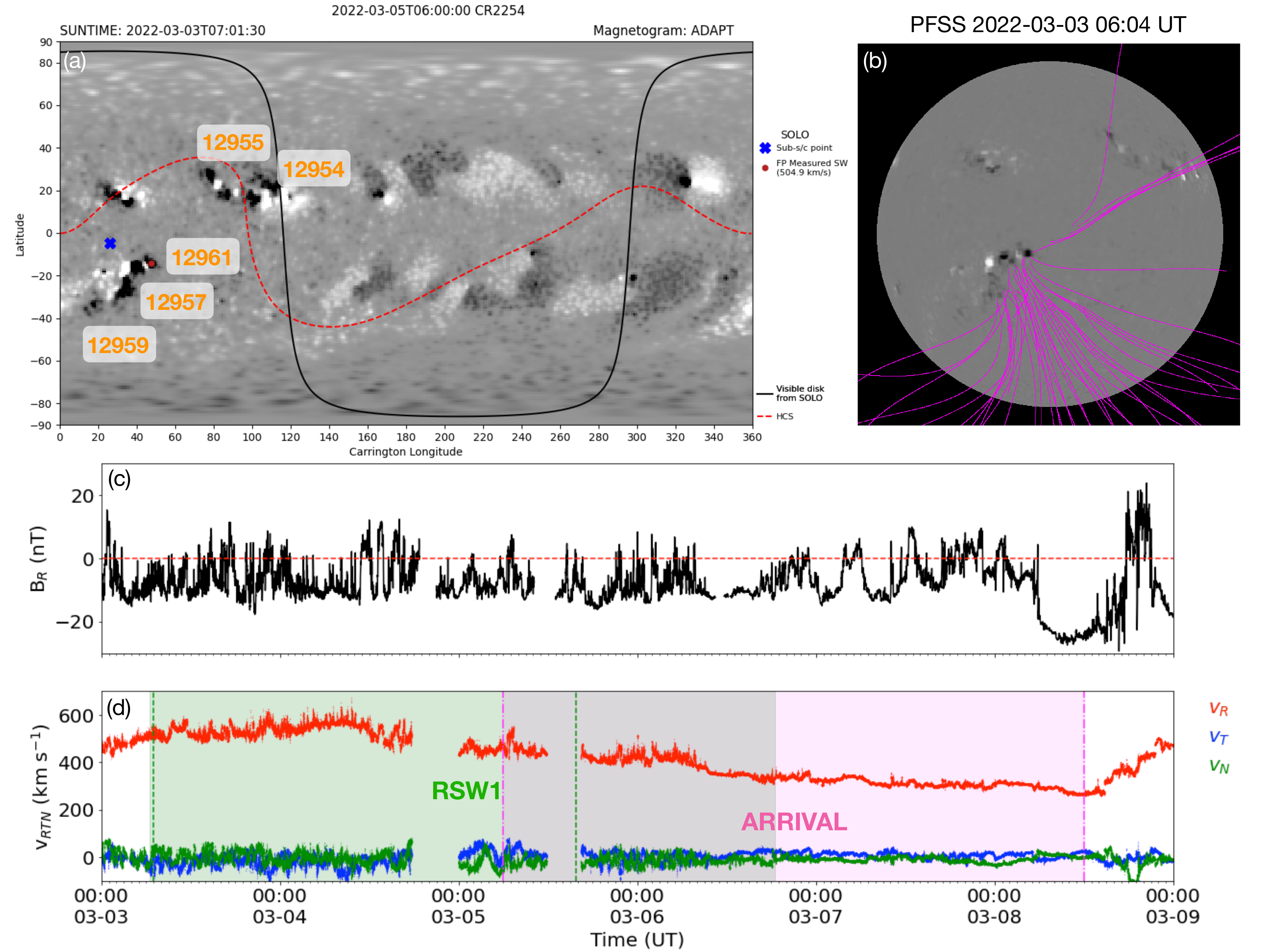}
\caption{The SWA/PAS and MAG data taken by Solar Orbiter during the time period of 2022 March 3 and March 9 along with the results from the connectivity tool and the PFSS model taken at the start of RSW1. Panel~a shows the ADAPT magnetic field map from 2022 March 3 at 07:01:30~UT used to calculate the connectivity of Solar Orbiter on 2022 March 6 at 06:00~UT. The connectivity points given by the red circles are determined using a solar  wind speed of 505~km~s$^{-1}$ measured by SWA/PAS. NOAA AR numbers are labelled in orange. The online animation of panel~a shows a movie of the magnetic connectivity between 2022 February 28 18:00~UT until March 8 18:00~UT (in situ time) with a cadence of 6~hr. Panel~b shows the PFSS model at a similar time at the start of RSW1. Black (white) represent negative (positive) magnetic field polarities with a saturation of $\pm$500~G. The magenta lines are representative open negative magnetic field lines located in the equatorial coronal hole and at the boundary of the AR complex (ARs 12957, 12959, 12961). Panel~c shows the radial component of the magnetic field measured by MAG where the red dotted line indicates 0~nT. Panel~d shows the SWA/PAS 3D velocity distribution (v$_{R}$, v$_{T}$, v$_{N}$). The green shaded region indicates the time period of RSW1 whereas the pink shows the corresponding solar wind arrival period taken from the connectivity tool. The first and last solar wind arrival (pink dashed dotted lines) correspond to the Sun time (green dashed lines) determined from the connectivity tool, when Solar Orbiter is connected to the AR complex i.e. the target of RSW1. \label{fig:fig9}} 
\end{figure*}

\section{Preliminary Data Analysis}

\subsection{In situ Measurements during and post Remote Sensing Window 1}

Figure~\ref{fig:fig9} shows the post-observation connectivity of Solar Orbiter along with the corresponding potential field source surface (PFSS) model, the SWA/PAS and MAG data for the time period 2022 March 3 to 9. On February 28 (in situ time), prior to the beginning of RSW1 of the Slow Wind SOOP (2022 March 3 at 06:00~UT), Solar Orbiter was connected to the positive polarity of AR 12954 in the northern hemisphere. The MAG radial magnetic field indicates that Solar Orbiter crossed the HCS on 2022 March 1 at 02:00~UT, which is around 6-10 hours later than had been predicted by the connectivity tool. The spacecraft is then connected to the negative polarity of AR 12955, which is to the east of AR 12954. During March 2 and 3, the connectivity of Solar Orbiter transitions from the negative polarity of AR 12955 across the boundary of the equatorial coronal hole (Figure~\ref{fig:fig9}~b), to the negative polarities of the target ARs 12957 and 12961 in the southern hemisphere. The magnetic connectivity remains stable during the entire RSW1 of the Slow Wind SOOP until 2022 March 8, when the connectivity transitions to the negative polarity of AR 12960. These results are supported by the large amount of negative open magnetic field present in the PFSS model (Figure~\ref{fig:fig9}~b) and the inward-directed (negative) radial magnetic field measured by MAG (Figure~\ref{fig:fig9}~c). The radial field remains predominantly negative for the entire period shown in panel~c apart from late on March 8, shortly after the arrival an ICME.

Figure~\ref{fig:fig9}~d shows the 3D velocity in RTN coordinates measured by SWA/PAS. Prior to and during the start of RSW1, the solar wind detected by PAS is moderately fast ($\sim$500~km~s$^{-1}$) and increases to almost 600~km~s$^{-1}$ on March 4. The solar wind speed then gradually decreases throughout and after the RSW to reach $\sim$300~km~s$^{-1}$, typical of the slow solar wind.

Using the connectivity tool, solar wind that arrived at Solar Orbiter on March 5 at 06:00~UT (505~km~s$^{-1}$, first magenta dashed-dot line in Figure~\ref{fig:fig9}), originated from on March 3 at 07:01~UT (first green dashed line in Figure~\ref{fig:fig9}~d) from the target observed during RSW1 (ARs 12957 and 12961). The last possible arrival of solar wind from the target is on March 8 at 12:00~UT (309~km~s$^{-1}$, second magenta dashed-dot line), when the connectivity shifts to the negative polarity of AR 12960. This corresponds to a solar wind transit time of just under two and three days, respectively. Therefore, the in situ data along with the post-observation connectivity analysis suggests that the magnetic connectivity tool was successful in providing a prediction for the target selection in RSW1.

%Switchbacks etc
When comparing the radial component of the magnetic field (Figure~\ref{fig:fig9}~c) to the radial proton velocity (Figure~\ref{fig:fig9}~d), there are time periods, particularly between $\sim$March 3--6, where there are reversals in the magnetic field that coincide with fluctuations in the radial velocity. These magnetic field reversals, where the magnetic field and proton velocity are highly correlated, are likely to be switchbacks and highly Alfv{\'e}nic in nature \citep{bale2019, kasper2019, horbury2018, horbury2020}. These structures are key to identifying the source region of the solar wind and the processes involved in its release such as interchange reconnection \citep{crooker2004, baker2009, owens2013, owens2018, bale2019, rouillard2020b}. A more in-depth analysis of the in situ measurements taken during and post RSW1 is presented in Yardley et al. (2023, in prep). For a study on Alfv{\'e}nic solar wind detected during the entire first Solar Orbiter perihelion see \citet{damicis2023}.

\subsection{In situ Measurements during and post Remote Sensing Window 2}

\begin{figure*}[ht!]
\plotone{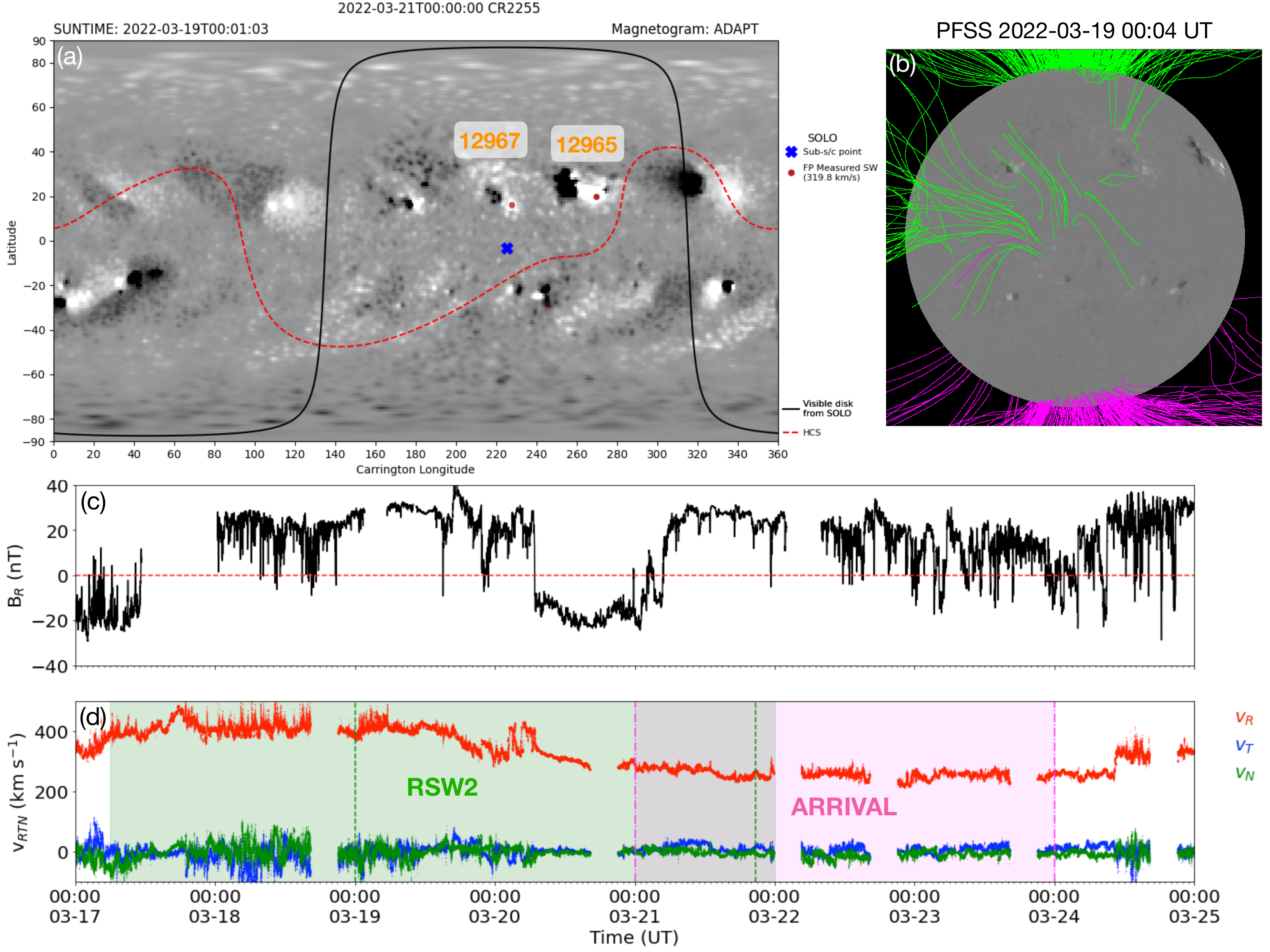}
\caption{The SWA/PAS and MAG data taken by Solar Orbiter during the time period of 2022 March 17 to 25, along with the connectivity tool and the PFSS model taken in the middle of RSW2. Panel~a shows the ADAPT magnetic field map from 2022 March 19 at 00:01:03~UT used to calculate the connectivity of Solar Orbiter on 2022 March 21 at 00:00~UT. The connectivity points (red circles) are determined by using a solar wind speed of 320~km~s$^{-1}$ measured by SWA/PAS. The online animation of panel~a shows a movie of the magnetic connectivity between 2022 March 16 00:00~UT until March 26 00:00~UT(in situ time) with a cadence of 6~hr. Panel~b shows an SDO/HMI magnetogram with open field lines taken from the PFSS model (created using the PFSS module from SSW developed by \citep[][]{schrijver2003}) taken at a similar time. Panel~c shows the radial magnetic field measured by MAG and panel~d shows the 3D velocity distribution functions measured by SWA/PAS. The green shaded region indicates the time period of RSW2 whereas the pink shows the corresponding solar wind arrival period taken from the connectivity tool. The first and last solar wind arrival (pink dash-dot lines) correspond to the Sun time (green dashed lines) determined from the connectivity tool, when Solar Orbiter is connected to the positive polarity of AR 12967 i.e. the second target of RSW2.  \label{fig:fig10}} 
\end{figure*}

Figure~\ref{fig:fig10} shows the magnetic connectivity tool prediction, along with the PFSS model on 2022 March 19, and the SWA/PAS and MAG data for the time period of 2022 March 17 until 25.

Prior to RSW2, Solar Orbiter was connected to the decayed negative magnetic field in the southern hemisphere, close to the west limb as viewed from Solar Orbiter. During March 16-17 there are multiple field reversals recorded in the radial magnetic field by MAG. The connectivity tool suggests that this is due to the connectivity transitioning between the decayed negative field, across the HCS, to the large positive polarity of AR 12965 (Figure~\ref{fig:fig10}~a). By 12:00~UT on 2022 March 18 (in situ time, corresponding to 21:26~UT on March 16 Suntime) the connectivity points for Solar Orbiter are situated entirely in the positive polarity of AR 12965. At the beginning of RSW2, on 2022 March 17, Solar Orbiter is observing the southern polar coronal hole boundary while the solar wind at Solar Orbiter, according to the connectivity tool, is arriving from AR 12965. On 2022 March 20 at 06:00~UT (corresponding to a Suntime of 2022 March 18 16:15~UT) the connectivity of Solar Orbiter changes to the positive polarity of decayed AR 12967. The positive polarity of this AR is observed by the remote sensing instruments on board Solar Orbiter from 2022 March 18 19:40~UT, shortly after the change in connectivity to this region, until the end of RSW2. Solar Orbiter is connected to the positive polarity of AR 12967 until March 26 at 12:00~UT (in situ time) when it transitions to decayed positive polarity field to the east of the AR. 

Around the time that the connectivity points begin to transition to the positive polarity of AR 12967 on March 19, there are two corridors of open positive magnetic field associated with this polarity (see panel b). There is one corridor that extends northwards towards the polar coronal hole and another extending south across to an equatorial hole. These regions appear dark in the EUV 193~\AA\ emission (see Figure~\ref{fig:fig6}). \citet{baker2023} explores the SWA, MAG, Hinode/EIS and SDO/AIA data, along with PFSS extrapolations and magnetic connectivity of Solar Orbiter to determine the corridor which is responsible for the slow solar wind that arrived at Solar Orbiter.  It was found that the extremely slow, high density solar wind detected by Solar Orbiter originates from the corridor north of AR 12967 and is characterised by moderate Alfv{\'e}nicity and switchback events.

Panel~c of Figure~\ref{fig:fig10} shows the radial magnetic field measured by MAG during March 17 to 25. It shows that after March 17 outward-directed (positive) radial magnetic field is measured for the entirety of the time period apart from when the field reverses briefly on March 20 due to the arrival of an ICME with negative B${_R}$. The 3D velocity distributions measured by SWA/PAS for the corresponding time period are shown in panel d. Initially, solar wind speeds of $\sim$400~km~s$^{-1}$ are measured, which gradually decreases to a very low speed of $\sim$210~km~s$^{-1}$ on 2022 March 22 and remains low for the next few days. This may be close to the slowest solar wind speed ever observed however, care should be taken when analysing solar wind with measured velocities below ~300~km~s$^{-1}$ due to the velocity distribution falling into the low energy range of PAS\footnote{\url{https://drive.google.com/drive/folders/1d2Y-G0BiAyAyQqTXL6x9zI39PosoWay\_}}.

The connectivity tool suggests that Solar Orbiter is connected to the remote sensing target (positive polarity of AR 12967) from the beginning of March 19 (Suntime, first green dashed line in panel d) with the first arrival of slow solar wind (320~km~s$^{-1}$ from this region from March 21 at 00:00~UT (first pink dashed dot line). The last arrival of slow solar wind (282~km~s$^{-1}$) from this region during RSW2 at Solar Orbiter is on March 24 at 00:00~UT (second pink dashed dot line), originating from the AR at 20:43~UT on 2022 March 21 (second green dashed line).

\section{Summary and Discussion}

%slow wind SOOP goals
The Slow Solar Wind SOOP has been designed in order to understand the origins and formation mechanisms of the slow solar wind. The targets of the Slow Wind SOOP are open-closed magnetic field boundaries such as the edges of coronal holes and the peripheries of active regions where plasma can escape directly along open magnetic field or indirectly i.e. due to interchange reconnection. Solar Orbiter's extensive suite of instruments and unique orbit provides the first opportunity to trace solar wind plasma measured in situ back to its source region on the Sun using data from a single space-based observatory.

%RSW1 and RSW2 run times and distance
The SOOP operated for the first time during Solar Orbiter's first close perihelion passage in March 2022. High-resolution observations were taken during two remote sensing windows, which took place between 2022 March 3 06:00~UT and 2022 March 6 at 18:30~UT (RSW1), and 2022 March 17 06:00~UT until 2022 March 22 (RSW2). Solar Orbiter was at a Heliocentric distance of 0.55-0.51 and 0.38-0.34~au, respectively, during these time periods. 

The targets for the high-resolution observations during both RSWs were chosen roughly three days in advance during pointing decision meetings based upon the predictions provided by the connectivity tool, low latency Solar Orbiter observations and supporting space-based observatories such as SDO and STEREO-A. 

%RSW1
A target for RSW1 was easily identified prior to the commencement of the remote-sensing observations on 2022 March 3 as Solar Orbiter was predicted to be magnetically connected to a negative polarity of an AR complex in the southern hemisphere for the entire observation window. The predicted connectivity and its evolution was also very similar during the previous solar rotation. Therefore, high-resolution remote sensing observations were taken of the AR complex including ARs 12957 and 12961. Post observation, the results from the connectivity tool show that the connectivity of Solar Orbiter evolves as follows. Solar Orbiter was connected to the positive polarity of AR 12954 in the northern hemisphere prior to RSW1, crossed the HCS to the negative polarity of AR 12955, traced the equatorial coronal hole boundary, before being connected to AR 12961 then 12957 during the entirety of RSW1. The in situ-data from Solar Orbiter supports this analysis. Firstly, inward-directed (negative) magnetic field is measured by MAG throughout and after RSW1. Secondly, the range of solar wind speeds between 670 and 260~km~s$^{-1}$ measured by SWA/PAS. Initially the arrival of moderately fast solar wind is observed followed by a gradual decrease to speeds typical of slow solar wind. These results, along with the connectivity tool, suggest that the solar wind originates from the equatorial coronal hole arrived at Solar Orbiter first, followed by the boundary of AR 12961 then 12957. A more detailed analysis is required using additional datasets taken by Solar Orbiter during this period to confirm the connection to this region. In particular, it would be important to trace abundance measurements made by SWA/HIS to coronal abundance measurements made using the SPICE composition studies. Oxygen and carbon charge states measured by SWA/HIS can also help to characterise the origin of the solar wind (Yardley et al. 2023, in preparation).

%RSW2
Identifying a target for RSW2 was more difficult compared to RSW1. The connectivity of Solar Orbiter was not only different during the previous solar rotation but it also changed significantly prior to and during RSW2. This was mainly due to the increase in number of active regions on the solar disk during this time period and in particular, the large amount of flux emergence occurring in AR 12965. Significant flux emergence influences the ADAPT magnetic field maps and hence the predicted connectivity. Therefore, the change in connectivity made it very challenging to select a target during the pointing decision meetings for RSW2. Ultimately, two targets were chosen: the boundary of the southern polar coronal hole and the decayed positive polarity of AR 12967. Analysing the connectivity post observation suggests the following evolution in connectivity. Before RSW2 Solar Orbiter was connected to the decayed negative polarity in the south before crossing the HCS to the positive polarity of AR 12965. From 2022 March 19 onwards (suntime, corresponding to an in situ time of March 21 12:00~UT) Solar Orbiter was connected to the decayed positive polarity of AR 12967. The MAG and SWA/PAS data support these conclusions. 

The radial magnetic field measured by MAG was positive from midday on 2022 March 17 onwards, apart from the arrival of an ICME on 2022 March 20. The solar wind speed measured is relatively slow throughout the window with initial speeds averaging around 400~km~s$^{-1}$, gradually decreasing to a minimum of 210~km~s$^{-1}$. The decrease in speed could be due to the change in connectivity from the large positive polarity of AR 12965 to the small, decayed positive polarity of AR 12967 and the typical decreasing radial velcity profile in the ICME. The in situ data, connectivity, and upflows in Hinode/EIS data is investigated further in \citet{baker2023}. 

Unfortunately, for this period we do not have SWA/HIS data as HIS reached its thermal limit and was switched off on 2022 March 13. This will make it more challenging to characterise the solar wind detected in situ with Solar Orbiter. \citet{nawin2023} focuses on the first target during RSW2 where a coronal dimming associated with a filament eruption merges with the southern polar coronal hole boundary. The SO/EUI, Hinode/EIS, SDO/AIA and HMI data suggest that there is component reconnection occurring between the boundary of the coronal hole and the coronal dimming.

% Conclusion - successful run of SOOP - Jazz it up
In Summary, the Slow Wind SOOP operated for 3.5 and 5 days, respectively, during the first two remote sensing windows of Solar Orbiter's close perihelion passage in March 2022. The SOOP campaign was very successful, given that the targets for the RSWs must be decided in advance and the targets depend upon the predicted magnetic connectivity of the spacecraft during these time periods. During the RSWs Solar Orbiter observed both primary targets i.e. the periphery of an AR and the boundary of a CH, outlined in the Slow Wind SOOP. Complementary data of these targets were also obtained by IRIS and Hinode through coordinated observation campaigns. During and after both remote sensing windows slow solar wind was measured by SWA/PAS. The radial magnetic field measured by MAG along with the connectivity tool suggest that the slow wind detected in situ by Solar Orbiter originated from two out of three of our selected targets, namely the boundaries of the negative polarity of AR 12961 during RSW1 and the periphery of the decayed positive polarity of AR 12967. 

The next steps are to validate these results by combining the analysis of a wider range of both remote sensing and in situ data taken from Solar Orbiter during these time periods. As the Slow Wind SOOP was very successful it will operate again during the next perihelion passage in March and April 2023 (LTP11), where the knowledge and experience gained from operating the SOOP during the first close perihelion will be applied.\\

\section{}

S.L.Y. would like to thank STFC via the consolidated grant (STFC ST/V000497/1), NERC via the SWIMMR Aviation Risk Modelling (SWARM) project grant (NE/V002899/1), and the Donostia International Physics Center for their hospitality. D.M.L. is grateful to the Science Technology and Facilities Council for the award of an Ernest Rutherford Fellowship (ST/R003246/1). D.B. is funded under STFC consolidated grant number ST/S000240/1. The work of D.H.B. was performed under contract to the Naval Research Laboratory and was funded by the NASA Hinode program. L.M.G. would like to thank NERC via the SWIMMR Aviation Modelling Risk (SWARM) project (grant no. NE/V002899/1). V.P. acknowledges support from NASA contract NNG09FA40C (IRIS) and NASA grant no. 80NSSC21K0623. L.v.D.G. acknowledges the Hungarian National Research, Development and Innovation Office grant OTKA K-131508. S.P. S.P. acknowledges the funding by CNES through the MEDOC data and operations center. W.T.T. was funded under NASA Grant NNG06EB68C. Solar Orbiter is a space mission of international collaboration between ESA and NASA, operated by ESA. We are grateful to the ESA SOC and MOC teams for their support. The SO/EUI instrument was built by CSL, IAS, MPS, MSSL/UCL, PMOD/WRC, ROB, LCF/IO with funding from the Belgian Federal Science Policy Office (BELSPO/PRODEX PEA 4000134088); the Centre National d’Etudes Spatiales (CNES); the UK Space Agency (UKSA); the Bundesministerium f\"{u}r Wirtschaft und Energie (BMWi) through the Deutsches Zentrum f\"{u}r Luft- und Raumfahrt (DLR); and the Swiss Space Office (SSO). The ROB team thanks the Belgian Federal Science Policy Office (BELSPO) for the provision of financial support in the framework of the PRODEX Programme of the European Space Agency (ESA) under contract numbers 4000134088, 4000112292, 4000136424, and 4000134474. Solar Orbiter magnetometer operations are funded by the UK Space Agency (grant ST/T001062/1). T.H. is supported by STFC grant ST/X002098/1. The German contribution to SO/PHI is funded by the BMWi through DLR and by MPG central funds. The Spanish contribution is funded by FEDER/AEI/MCIU (RTI2018-096886-C5), a “Center of Excellence Severo Ochoa” award to IAA-CSIC (SEV-2017-0709), and a Ramón y Cajal fellowship awarded to DOS. The French contribution is funded by CNES. J.C.T.I. and D.O.S. are supported by the Spanish Ministry of Economy and Competitiveness through projects ESP-2016- 77548-C5-1-R, and by the Spanish Science Ministry “Centro de Excelencia Severo Ochoa” Program under grant SEV-2017- 0709 and project RTI2018-096886-B-C51. D.O.S. also acknowledges financial support from a Ramón y Cajal fellowship. The development of the SPICE instrument was funded by ESA and ESA member states (France, Germany, Norway, Switzerland, United Kingdom). The SPICE hardware consortium was led by Science and Technology Facilities Council (STFC) RAL Space and included Institut d’Astrophysique Spatiale (IAS), Max-Planck-Institut für Sonnensystemforschung (MPS), Physikalisch-Meteorologisches Observatorium Davos and World Radiation Center (PMOD/WRC), Institute of Theoretical Astrophysics (University of Oslo), NASA Goddard Space Flight Center (GSFC) and Southwest Research Institute (SwRI). The in-ﬂight commissioning of SPICE was led by the instrument team at UKRI/STFC RAL Space. AF’s and AG’s research is funded by UKRI STFC. SO/SWA data are derived from scientific sensors which have been designed and created, and are operated under funding provided in numerous contracts from the UK Space Agency (UKSA), the UK Science and Technology Facilities Council (STFC), the Agenzia Spaziale Italiana (ASI), the Centre National d’Etudes Spatiales (CNES,France), the Centre National de la Recherche Scientifique (CNRS, France), the Czech contribution to the ESA PRODEX programme and NASA. Solar Orbiter SWA work at UCL/MSSL was funded under STFC grants ST/T001356/1, ST/S000240/1, ST/X002152/1 and ST/W001004/1. SDO data are courtesy of NASA/SDO and the AIA, EVE, and HMI science teams. IRIS is a NASA small explorer mission developed and operated by LMSAL with mission operations executed at NASA Ames Research Center and major contributions to downlink communications funded by ESA and the Norwegian Space Centre. CHIANTI is a collaborative project involving George Mason University, the University of Michigan (USA), University of Cambridge (UK) and NASA Goddard Space Flight Center (USA). This work also utilises data produced collaboratively between AFRL/ADAPT and NSO/NISP. For the purpose of open access, the author has applied a ‘Creative Commons Attribution (CC BY) licence (where permitted by UKRI, ‘Open Government Licence’ or ‘Creative Commons Attribution No-derivatives (CC BY ND) licence’ may be stated instead) to any Author Accepted Manuscript version arising.

Facilities: Solar Orbiter (SO), Solar Dynamics Observatory (SDO), Hinode, Interface Region Imaging Spectrograph (IRIS). 

Data Availability: All data published in this manuscript are publicly available online. Solar Orbiter data is available via the SOAR archive (\url{https://soar.esac.esa.int/soar/}), data from the Solar Dynamics Observatory can be downloaded from JSOC (\url{http://jsoc.stanford.edu/}), finally data taken by Hinode and IRIS can be retrieved from the OSLO database (\url{http://sdc.uio.no/sdc/}).

Software: JHelioviewer \citep{muller2017}, SunPy \citep{sunpy2020} v4.1.0 (DOI:10.5281/zenodo.7314636), AstroPy \citep{astropy2022}, SolarSoft IDL PFSS package \citep{schrijver2003}.

%\end{acknowledgments}

\bibliography{thebibliography}{}

\begin{thebibliography}{}
\expandafter\ifx\csname natexlab\endcsname\relax\def\natexlab#1{#1}\fi
\providecommand{\url}[1]{\href{#1}{#1}}
\providecommand{\dodoi}[1]{doi:~\href{http://doi.org/#1}{\nolinkurl{#1}}}
\providecommand{\doeprint}[1]{\href{http://ascl.net/#1}{\nolinkurl{http://ascl.net/#1}}}
\providecommand{\doarXiv}[1]{\href{https://arxiv.org/abs/#1}{\nolinkurl{https://arxiv.org/abs/#1}}}

\bibitem[{{Abbo} {et~al.}(2016){Abbo}, {Ofman}, {Antiochos}, {Hansteen},
  {Harra}, {Ko}, {Lapenta}, {Li}, {Riley}, {Strachan}, {von Steiger}, \&
  {Wang}}]{abbo2016}
{Abbo}, L., {Ofman}, L., {Antiochos}, S.~K., {et~al.} 2016, \ssr, 201, 55,
  \dodoi{10.1007/s11214-016-0264-1}

\bibitem[{{Antiochos} {et~al.}(2011){Antiochos}, {Miki{\'c}}, {Titov},
  {Lionello}, \& {Linker}}]{antiochos2011}
{Antiochos}, S.~K., {Miki{\'c}}, Z., {Titov}, V.~S., {Lionello}, R., \&
  {Linker}, J.~A. 2011, \apj, 731, 112, \dodoi{10.1088/0004-637X/731/2/112}

\bibitem[{{Antonucci} {et~al.}(2020){Antonucci}, {Romoli}, {Andretta},
  {Fineschi}, {Heinzel}, {Moses}, {Naletto}, {Nicolini}, {Spadaro}, {Teriaca},
  {Berlicki}, {Capobianco}, {Crescenzio}, {Da Deppo}, {Focardi}, {Frassetto},
  {Heerlein}, {Landini}, {Magli}, {Marco Malvezzi}, {Massone}, {Melich},
  {Nicolosi}, {Noci}, {Pancrazzi}, {Pelizzo}, {Poletto}, {Sasso},
  {Sch{\"u}hle}, {Solanki}, {Strachan}, {Susino}, {Tondello}, {Uslenghi},
  {Woch}, {Abbo}, {Bemporad}, {Casti}, {Dolei}, {Grimani}, {Messerotti},
  {Ricci}, {Straus}, {Telloni}, {Zuppella}, {Auch{\`e}re}, {Bruno},
  {Ciaravella}, {Corso}, {Alvarez Copano}, {Aznar Cuadrado}, {D'Amicis},
  {Enge}, {Gravina}, {Jej{\v{c}}i{\v{c}}}, {Lamy}, {Lanzafame}, {Meierdierks},
  {Papagiannaki}, {Peter}, {Fernandez Rico}, {Giday Sertsu}, {Staub},
  {Tsinganos}, {Velli}, {Ventura}, {Verroi}, {Vial}, {Vives}, {Volpicelli},
  {Werner}, {Zerr}, {Negri}, {Castronuovo}, {Gabrielli}, {Bertacin},
  {Carpentiero}, {Natalucci}, {Marliani}, {Cesa}, {Laget}, {Morea},
  {Pieraccini}, {Radaelli}, {Sandri}, {Sarra}, {Cesare}, {Del Forno}, {Massa},
  {Montabone}, {Mottini}, {Quattropani}, {Schillaci}, {Boccardo}, {Brando},
  {Pandi}, {Baietto}, {Bertone}, {Alvarez-Herrero}, {Garc{\'\i}a Parejo},
  {Cebollero}, {Amoruso}, \& {Centonze}}]{antonucci2020}
{Antonucci}, E., {Romoli}, M., {Andretta}, V., {et~al.} 2020, \aap, 642, A10,
  \dodoi{10.1051/0004-6361/201935338}

\bibitem[{{Arge} {et~al.}(2010){Arge}, {Henney}, {Koller}, {Compeau}, {Young},
  {MacKenzie}, {Fay}, \& {Harvey}}]{arge2010}
{Arge}, C.~N., {Henney}, C.~J., {Koller}, J., {et~al.} 2010, in American
  Institute of Physics Conference Series, Vol. 1216, Twelfth International
  Solar Wind Conference, ed. M.~{Maksimovic}, K.~{Issautier},
  N.~{Meyer-Vernet}, M.~{Moncuquet}, \& F.~{Pantellini}, 343--346,
  \dodoi{10.1063/1.3395870}

\bibitem[{{Astropy Collaboration} {et~al.}(2022){Astropy Collaboration},
  {Price-Whelan}, {Lim}, {Earl}, {Starkman}, {Bradley}, {Shupe}, {Patil},
  {Corrales}, {Brasseur}, {N{\"o}the}, {Donath}, {Tollerud}, {Morris},
  {Ginsburg}, {Vaher}, {Weaver}, {Tocknell}, {Jamieson}, {van Kerkwijk},
  {Robitaille}, {Merry}, {Bachetti}, {G{\"u}nther}, {Aldcroft},
  {Alvarado-Montes}, {Archibald}, {B{\'o}di}, {Bapat}, {Barentsen},
  {Baz{\'a}n}, {Biswas}, {Boquien}, {Burke}, {Cara}, {Cara}, {Conroy},
  {Conseil}, {Craig}, {Cross}, {Cruz}, {D'Eugenio}, {Dencheva}, {Devillepoix},
  {Dietrich}, {Eigenbrot}, {Erben}, {Ferreira}, {Foreman-Mackey}, {Fox},
  {Freij}, {Garg}, {Geda}, {Glattly}, {Gondhalekar}, {Gordon}, {Grant},
  {Greenfield}, {Groener}, {Guest}, {Gurovich}, {Handberg}, {Hart},
  {Hatfield-Dodds}, {Homeier}, {Hosseinzadeh}, {Jenness}, {Jones}, {Joseph},
  {Kalmbach}, {Karamehmetoglu}, {Ka{\l}uszy{\'n}ski}, {Kelley}, {Kern},
  {Kerzendorf}, {Koch}, {Kulumani}, {Lee}, {Ly}, {Ma}, {MacBride}, {Maljaars},
  {Muna}, {Murphy}, {Norman}, {O'Steen}, {Oman}, {Pacifici}, {Pascual},
  {Pascual-Granado}, {Patil}, {Perren}, {Pickering}, {Rastogi}, {Roulston},
  {Ryan}, {Rykoff}, {Sabater}, {Sakurikar}, {Salgado}, {Sanghi}, {Saunders},
  {Savchenko}, {Schwardt}, {Seifert-Eckert}, {Shih}, {Jain}, {Shukla}, {Sick},
  {Simpson}, {Singanamalla}, {Singer}, {Singhal}, {Sinha}, {Sip{\H{o}}cz},
  {Spitler}, {Stansby}, {Streicher}, {{\v{S}}umak}, {Swinbank}, {Taranu},
  {Tewary}, {Tremblay}, {Val-Borro}, {Van Kooten}, {Vasovi{\'c}}, {Verma}, {de
  Miranda Cardoso}, {Williams}, {Wilson}, {Winkel}, {Wood-Vasey}, {Xue},
  {Yoachim}, {Zhang}, {Zonca}, \& {Astropy Project Contributors}}]{astropy2022}
{Astropy Collaboration}, {Price-Whelan}, A.~M., {Lim}, P.~L., {et~al.} 2022,
  \apj, 935, 167, \dodoi{10.3847/1538-4357/ac7c74}

\bibitem[{{Auch{\`e}re} {et~al.}(2020){Auch{\`e}re}, {Andretta}, {Antonucci},
  {Bach}, {Battaglia}, {Bemporad}, {Berghmans}, {Buchlin}, {Caminade},
  {Carlsson}, {Carlyle}, {Cerullo}, {Chamberlin}, {Colaninno}, {Davila}, {De
  Groof}, {Etesi}, {Fahmy}, {Fineschi}, {Fludra}, {Gilbert}, {Giunta},
  {Grundy}, {Haberreiter}, {Harra}, {Hassler}, {Hirzberger}, {Howard},
  {Hurford}, {Kleint}, {Kolleck}, {Krucker}, {Lagg}, {Landini}, {Long},
  {Lefort}, {Lodiot}, {Mampaey}, {Maloney}, {Marliani}, {Martinez-Pillet},
  {McMullin}, {M{\"u}ller}, {Nicolini}, {Orozco Suarez}, {Pacros}, {Pancrazzi},
  {Parenti}, {Peter}, {Philippon}, {Plunkett}, {Rich}, {Rochus}, {Rouillard},
  {Romoli}, {Sanchez}, {Sch{\"u}hle}, {Sidher}, {Solanki}, {Spadaro}, {St Cyr},
  {Straus}, {Tanco}, {Teriaca}, {Thompson}, {del Toro Iniesta}, {Verbeeck},
  {Vourlidas}, {Watson}, {Wiegelmann}, {Williams}, {Woch}, {Zhukov}, \&
  {Zouganelis}}]{auchere2020}
{Auch{\`e}re}, F., {Andretta}, V., {Antonucci}, E., {et~al.} 2020, \aap, 642,
  A6, \dodoi{10.1051/0004-6361/201937032}

\bibitem[{{Baker} {et~al.}(2017){Baker}, {Janvier}, {D{\'e}moulin}, \&
  {Mandrini}}]{baker2017}
{Baker}, D., {Janvier}, M., {D{\'e}moulin}, P., \& {Mandrini}, C.~H. 2017,
  \solphys, 292, 46, \dodoi{10.1007/s11207-017-1072-9}

\bibitem[{{Baker} {et~al.}(2009){Baker}, {van Driel-Gesztelyi}, {Mandrini},
  {D{\'e}moulin}, \& {Murray}}]{baker2009}
{Baker}, D., {van Driel-Gesztelyi}, L., {Mandrini}, C.~H., {D{\'e}moulin}, P.,
  \& {Murray}, M.~J. 2009, \apj, 705, 926, \dodoi{10.1088/0004-637X/705/1/926}

\bibitem[{{Baker} {et~al.}(2023){Baker}, {Demoulin}, {Yardley}, {Mihailescu},
  {van Driel-Gesztelyi}, {D'Amicis}, {Long}, {To}, {Owen}, {Horbury}, {Brooks},
  {Perrone}, {French}, {James}, {Janvier}, {Matthews}, {Stangalini}, {Valori},
  {Smith}, {Anzar Cuadrado}, {Peter}, {Schuehle}, {Harra}, {Barczynski},
  {Berghmans}, {Zhukov}, {Rodriguez}, \& {Verbeeck}}]{baker2023}
{Baker}, D., {Demoulin}, P., {Yardley}, S.~L., {et~al.} 2023, arXiv e-prints,
  arXiv:2303.12192, \dodoi{10.48550/arXiv.2303.12192}

\bibitem[{{Bale} {et~al.}(2019){Bale}, {Badman}, {Bonnell}, {Bowen}, {Burgess},
  {Case}, {Cattell}, {Chandran}, {Chaston}, {Chen}, {Drake}, {de Wit},
  {Eastwood}, {Ergun}, {Farrell}, {Fong}, {Goetz}, {Goldstein}, {Goodrich},
  {Harvey}, {Horbury}, {Howes}, {Kasper}, {Kellogg}, {Klimchuk}, {Korreck},
  {Krasnoselskikh}, {Krucker}, {Laker}, {Larson}, {MacDowall}, {Maksimovic},
  {Malaspina}, {Martinez-Oliveros}, {McComas}, {Meyer-Vernet}, {Moncuquet},
  {Mozer}, {Phan}, {Pulupa}, {Raouafi}, {Salem}, {Stansby}, {Stevens}, {Szabo},
  {Velli}, {Woolley}, \& {Wygant}}]{bale2019}
{Bale}, S.~D., {Badman}, S.~T., {Bonnell}, J.~W., {et~al.} 2019, \nat, 576,
  237, \dodoi{10.1038/s41586-019-1818-7}

\bibitem[{{Berghmans} {et~al.}(2023){Berghmans}, {Antolin}, {Auch{\`e}re},
  {Aznar Cuadrado}, {Barczynski}, {Chitta}, {Gissot}, {Harra}, {Huang},
  {Janvier}, {Kraaikamp}, {Long}, {Mandal}, {Mierla}, {Parenti}, {Peter},
  {Rodriguez}, {Sch{\"u}hle}, {Smith}, {Solanki}, {Stegen}, {Teriaca},
  {Verbeeck}, {West}, {Zhukov}, {Appourchaux}, {Aulanier}, {Buchlin},
  {Delmotte}, {Gilles}, {Haberreiter}, {Halain}, {Heerlein}, {Hochedez}, {Gyo},
  {Poedts}, \& {Rochus}}]{berghmans2023}
{Berghmans}, D., {Antolin}, P., {Auch{\`e}re}, F., {et~al.} 2023, arXiv
  e-prints, arXiv:2301.05616, \dodoi{10.48550/arXiv.2301.05616}

\bibitem[{{Brooks} {et~al.}(2022){Brooks}, {Baker}, {van Driel-Gesztelyi},
  {Warren}, \& {Yardley}}]{brooks2022}
{Brooks}, D.~H., {Baker}, D., {van Driel-Gesztelyi}, L., {Warren}, H.~P., \&
  {Yardley}, S.~L. 2022, \apjl, 930, L10, \dodoi{10.3847/2041-8213/ac6878}

\bibitem[{{Brooks} {et~al.}(2021){Brooks}, {Harra}, {Bale}, {Barczynski},
  {Mandrini}, {Polito}, \& {Warren}}]{brooks2021a}
{Brooks}, D.~H., {Harra}, L., {Bale}, S.~D., {et~al.} 2021, \apj, 917, 25,
  \dodoi{10.3847/1538-4357/ac0917}

\bibitem[{{Brooks} {et~al.}(2015){Brooks}, {Ugarte-Urra}, \&
  {Warren}}]{brooks2015}
{Brooks}, D.~H., {Ugarte-Urra}, I., \& {Warren}, H.~P. 2015, Nature
  Communications, 6, 5947, \dodoi{10.1038/ncomms6947}

\bibitem[{{Brooks} \& {Warren}(2011)}]{brooks2011}
{Brooks}, D.~H., \& {Warren}, H.~P. 2011, \apjl, 727, L13,
  \dodoi{10.1088/2041-8205/727/1/L13}

\bibitem[{{Brooks} \& {Yardley}(2021)}]{brooks2021b}
{Brooks}, D.~H., \& {Yardley}, S.~L. 2021, \mnras, 508, 1831,
  \dodoi{10.1093/mnras/stab2681}

\bibitem[{{Cranmer}(2009)}]{cranmer2009}
{Cranmer}, S.~R. 2009, \apj, 706, 824, \dodoi{10.1088/0004-637X/706/1/824}

\bibitem[{{Cranmer} {et~al.}(2017){Cranmer}, {Gibson}, \&
  {Riley}}]{cranmer2017}
{Cranmer}, S.~R., {Gibson}, S.~E., \& {Riley}, P. 2017, \ssr, 212, 1345,
  \dodoi{10.1007/s11214-017-0416-y}

\bibitem[{{Crooker} {et~al.}(2004){Crooker}, {Kahler}, {Larson}, \&
  {Lin}}]{crooker2004}
{Crooker}, N.~U., {Kahler}, S.~W., {Larson}, D.~E., \& {Lin}, R.~P. 2004,
  Journal of Geophysical Research (Space Physics), 109, A03108,
  \dodoi{10.1029/2003JA010278}

\bibitem[{{Crooker} \& {Owens}(2012)}]{crooker2012}
{Crooker}, N.~U., \& {Owens}, M.~J. 2012, \ssr, 172, 201,
  \dodoi{10.1007/s11214-011-9748-1}

\bibitem[{{Culhane} {et~al.}(2007){Culhane}, {Harra}, {James}, {Al-Janabi},
  {Bradley}, {Chaudry}, {Rees}, {Tandy}, {Thomas}, {Whillock}, {Winter},
  {Doschek}, {Korendyke}, {Brown}, {Myers}, {Mariska}, {Seely}, {Lang}, {Kent},
  {Shaughnessy}, {Young}, {Simnett}, {Castelli}, {Mahmoud}, {Mapson-Menard},
  {Probyn}, {Thomas}, {Davila}, {Dere}, {Windt}, {Shea}, {Hagood}, {Moye},
  {Hara}, {Watanabe}, {Matsuzaki}, {Kosugi}, {Hansteen}, \&
  {Wikstol}}]{Culhane2007}
{Culhane}, J.~L., {Harra}, L.~K., {James}, A.~M., {et~al.} 2007, \solphys, 243,
  19, \dodoi{10.1007/s01007-007-0293-1}

\bibitem[{{D'Amicis} \& {Bruno}(2015)}]{damicis2015}
{D'Amicis}, R., \& {Bruno}, R. 2015, \apj, 805, 84,
  \dodoi{10.1088/0004-637X/805/1/84}

\bibitem[{{D'Amicis} \& {et al.}(2023)}]{damicis2023}
{D'Amicis}, R., \& {et al.} 2023, \apj, in prep.

\bibitem[{{D'Amicis} {et~al.}(2019){D'Amicis}, {Matteini}, \&
  {Bruno}}]{damicis2019}
{D'Amicis}, R., {Matteini}, L., \& {Bruno}, R. 2019, \mnras, 483, 4665,
  \dodoi{10.1093/mnras/sty3329}

\bibitem[{{De Pontieu} {et~al.}(2014){De Pontieu}, {Title}, {Lemen}, {Kushner},
  {Akin}, {Allard}, {Berger}, {Boerner}, {Cheung}, {Chou}, {Drake}, {Duncan},
  {Freeland}, {Heyman}, {Hoffman}, {Hurlburt}, {Lindgren}, {Mathur}, {Rehse},
  {Sabolish}, {Seguin}, {Schrijver}, {Tarbell}, {W{\"u}lser}, {Wolfson},
  {Yanari}, {Mudge}, {Nguyen-Phuc}, {Timmons}, {van Bezooijen}, {Weingrod},
  {Brookner}, {Butcher}, {Dougherty}, {Eder}, {Knagenhjelm}, {Larsen},
  {Mansir}, {Phan}, {Boyle}, {Cheimets}, {DeLuca}, {Golub}, {Gates}, {Hertz},
  {McKillop}, {Park}, {Perry}, {Podgorski}, {Reeves}, {Saar}, {Testa}, {Tian},
  {Weber}, {Dunn}, {Eccles}, {Jaeggli}, {Kankelborg}, {Mashburn}, {Pust},
  {Springer}, {Carvalho}, {Kleint}, {Marmie}, {Mazmanian}, {Pereira}, {Sawyer},
  {Strong}, {Worden}, {Carlsson}, {Hansteen}, {Leenaarts}, {Wiesmann},
  {Aloise}, {Chu}, {Bush}, {Scherrer}, {Brekke}, {Martinez-Sykora}, {Lites},
  {McIntosh}, {Uitenbroek}, {Okamoto}, {Gummin}, {Auker}, {Jerram}, {Pool}, \&
  {Waltham}}]{depontieu2014}
{De Pontieu}, B., {Title}, A.~M., {Lemen}, J.~R., {et~al.} 2014, \solphys, 289,
  2733, \dodoi{10.1007/s11207-014-0485-y}

\bibitem[{{De Pontieu} {et~al.}(2021){De Pontieu}, {Polito}, {Hansteen},
  {Testa}, {Reeves}, {Antolin}, {N{\'o}brega-Siverio}, {Kowalski},
  {Martinez-Sykora}, {Carlsson}, {McIntosh}, {Liu}, {Daw}, \&
  {Kankelborg}}]{depontieu2021}
{De Pontieu}, B., {Polito}, V., {Hansteen}, V., {et~al.} 2021, \solphys, 296,
  84, \dodoi{10.1007/s11207-021-01826-0}

\bibitem[{{Del Zanna} {et~al.}(2021){Del Zanna}, {Dere}, {Young}, \&
  {Landi}}]{delzanna2021}
{Del Zanna}, G., {Dere}, K.~P., {Young}, P.~R., \& {Landi}, E. 2021, \apj, 909,
  38, \dodoi{10.3847/1538-4357/abd8ce}

\bibitem[{{D{\'e}moulin} {et~al.}(2013){D{\'e}moulin}, {Baker}, {Mandrini}, \&
  {van Driel-Gesztelyi}}]{demoulin2013}
{D{\'e}moulin}, P., {Baker}, D., {Mandrini}, C.~H., \& {van Driel-Gesztelyi},
  L. 2013, \solphys, 283, 341, \dodoi{10.1007/s11207-013-0234-7}

\bibitem[{{Dere} {et~al.}(1997){Dere}, {Landi}, {Mason}, {Monsignori Fossi}, \&
  {Young}}]{dere1997}
{Dere}, K.~P., {Landi}, E., {Mason}, H.~E., {Monsignori Fossi}, B.~C., \&
  {Young}, P.~R. 1997, \aaps, 125, 149, \dodoi{10.1051/aas:1997368}

\bibitem[{{Edwards} {et~al.}(2016){Edwards}, {Parnell}, {Harra}, {Culhane}, \&
  {Brooks}}]{edwards2016}
{Edwards}, S.~J., {Parnell}, C.~E., {Harra}, L.~K., {Culhane}, J.~L., \&
  {Brooks}, D.~H. 2016, \solphys, 291, 117, \dodoi{10.1007/s11207-015-0807-8}

\bibitem[{{Fisk} \& {Schwadron}(2001)}]{fisk2001}
{Fisk}, L.~A., \& {Schwadron}, N.~A. 2001, \apj, 560, 425,
  \dodoi{10.1086/322503}

\bibitem[{{Fludra} {et~al.}(2021){Fludra}, {Caldwell}, {Giunta}, {Grundy},
  {Guest}, {Leeks}, {Sidher}, {Auch{\`e}re}, {Carlsson}, {Hassler}, {Peter},
  {Aznar Cuadrado}, {Buchlin}, {Caminade}, {DeForest}, {Fredvik},
  {Haberreiter}, {Harra}, {Janvier}, {Kucera}, {M{\"u}ller}, {Parenti},
  {Schmutz}, {Sch{\"u}hle}, {Solanki}, {Teriaca}, {Thompson}, {Tustain},
  {Williams}, {Young}, \& {Chitta}}]{fludra2021}
{Fludra}, A., {Caldwell}, M., {Giunta}, A., {et~al.} 2021, \aap, 656, A38,
  \dodoi{10.1051/0004-6361/202141221}

\bibitem[{{Gandorfer} {et~al.}(2018){Gandorfer}, {Grauf}, {Staub}, {Bischoff},
  {Woch}, {Hirzberger}, {Solanki}, {{\'A}lvarez-Herrero}, {Garc{\'\i}a Parejo},
  {Schmidt}, {Volkmer}, {Appourchaux}, \& {del Toro Iniesta}}]{gandorfer2018}
{Gandorfer}, A., {Grauf}, B., {Staub}, J., {et~al.} 2018, in Society of
  Photo-Optical Instrumentation Engineers (SPIE) Conference Series, Vol. 10698,
  Space Telescopes and Instrumentation 2018: Optical, Infrared, and Millimeter
  Wave, ed. M.~{Lystrup}, H.~A. {MacEwen}, G.~G. {Fazio}, N.~{Batalha},
  N.~{Siegler}, \& E.~C. {Tong}, 106984N, \dodoi{10.1117/12.2311816}

\bibitem[{{Garc{\'\i}a Marirrodriga} {et~al.}(2021){Garc{\'\i}a Marirrodriga},
  {Pacros}, {Strandmoe}, {Arcioni}, {Arts}, {Ashcroft}, {Ayache}, {Bonnefous},
  {Brahimi}, {Cipriani}, {Damasio}, {De Jong}, {D{\'e}prez}, {Fahmy}, {Fels},
  {Fiebrich}, {Hass}, {Hern{\'a}ndez}, {Icardi}, {Junge}, {Kletzkine}, {Laget},
  {Le Deuff}, {Liebold}, {Lodiot}, {Marliani}, {Mascarello}, {M{\"u}ller},
  {Oganessian}, {Olivier}, {Palombo}, {Philippe}, {Ragnit}, {Ramachandran},
  {S{\'a}nchez P{\'e}rez}, {Stienstra}, {Th{\"u}rey}, {Urwin}, {Wirth}, \&
  {Zouganelis}}]{garcia2021}
{Garc{\'\i}a Marirrodriga}, C., {Pacros}, A., {Strandmoe}, S., {et~al.} 2021,
  \aap, 646, A121, \dodoi{10.1051/0004-6361/202038519}

\bibitem[{{Golub} {et~al.}(2007){Golub}, {Deluca}, {Austin}, {Bookbinder},
  {Caldwell}, {Cheimets}, {Cirtain}, {Cosmo}, {Reid}, {Sette}, {Weber},
  {Sakao}, {Kano}, {Shibasaki}, {Hara}, {Tsuneta}, {Kumagai}, {Tamura},
  {Shimojo}, {McCracken}, {Carpenter}, {Haight}, {Siler}, {Wright}, {Tucker},
  {Rutledge}, {Barbera}, {Peres}, \& {Varisco}}]{golub2007}
{Golub}, L., {Deluca}, E., {Austin}, G., {et~al.} 2007, \solphys, 243, 63,
  \dodoi{10.1007/s11207-007-0182-1}

\bibitem[{{Harra} {et~al.}(2008){Harra}, {Sakao}, {Mandrini}, {Hara}, {Imada},
  {Young}, {van Driel-Gesztelyi}, \& {Baker}}]{harra2008}
{Harra}, L.~K., {Sakao}, T., {Mandrini}, C.~H., {et~al.} 2008, \apjl, 676,
  L147, \dodoi{10.1086/587485}

\bibitem[{{Hickmann} {et~al.}(2015){Hickmann}, {Godinez}, {Henney}, \&
  {Arge}}]{hickmann2015}
{Hickmann}, K.~S., {Godinez}, H.~C., {Henney}, C.~J., \& {Arge}, C.~N. 2015,
  \solphys, 290, 1105, \dodoi{10.1007/s11207-015-0666-3}

\bibitem[{{Hinode Review Team} {et~al.}(2019){Hinode Review Team}, {Al-Janabi},
  {Antolin}, {Baker}, {Bellot Rubio}, {Bradley}, {Brooks}, {Centeno},
  {Culhane}, {Del Zanna}, {Doschek}, {Fletcher}, {Hara}, {Harra}, {Hillier},
  {Imada}, {Klimchuk}, {Mariska}, {Pereira}, {Reeves}, {Sakao}, {Sakurai},
  {Shimizu}, {Shimojo}, {Shiota}, {Solanki}, {Sterling}, {Su}, {Suematsu},
  {Tarbell}, {Tiwari}, {Toriumi}, {Ugarte-Urra}, {Warren}, {Watanabe}, \&
  {Young}}]{hinode2019}
{Hinode Review Team}, {Al-Janabi}, K., {Antolin}, P., {et~al.} 2019, \pasj, 71,
  R1, \dodoi{10.1093/pasj/psz084}

\bibitem[{{Horbury} {et~al.}(2018){Horbury}, {Matteini}, \&
  {Stansby}}]{horbury2018}
{Horbury}, T.~S., {Matteini}, L., \& {Stansby}, D. 2018, \mnras, 478, 1980,
  \dodoi{10.1093/mnras/sty953}

\bibitem[{{Horbury} {et~al.}(2020){Horbury}, {O'Brien}, {Carrasco Blazquez},
  {Bendyk}, {Brown}, {Hudson}, {Evans}, {Oddy}, {Carr}, {Beek}, {Cupido},
  {Bhattacharya}, {Dominguez}, {Matthews}, {Myklebust}, {Whiteside}, {Bale},
  {Baumjohann}, {Burgess}, {Carbone}, {Cargill}, {Eastwood}, {Erd{\"o}s},
  {Fletcher}, {Forsyth}, {Giacalone}, {Glassmeier}, {Goldstein}, {Hoeksema},
  {Lockwood}, {Magnes}, {Maksimovic}, {Marsch}, {Matthaeus}, {Murphy},
  {Nakariakov}, {Owen}, {Owens}, {Rodriguez-Pacheco}, {Richter}, {Riley},
  {Russell}, {Schwartz}, {Vainio}, {Velli}, {Vennerstrom}, {Walsh},
  {Wimmer-Schweingruber}, {Zank}, {M{\"u}ller}, {Zouganelis}, \&
  {Walsh}}]{horbury2020}
{Horbury}, T.~S., {O'Brien}, H., {Carrasco Blazquez}, I., {et~al.} 2020, \aap,
  642, A9, \dodoi{10.1051/0004-6361/201937257}

\bibitem[{{Howard} {et~al.}(2020){Howard}, {Vourlidas}, {Colaninno},
  {Korendyke}, {Plunkett}, {Carter}, {Wang}, {Rich}, {Lynch}, {Thurn},
  {Socker}, {Thernisien}, {Chua}, {Linton}, {Koss}, {Tun-Beltran}, {Dennison},
  {Stenborg}, {McMullin}, {Hunt}, {Baugh}, {Clifford}, {Keller}, {Janesick},
  {Tower}, {Grygon}, {Farkas}, {Hagood}, {Eisenhauer}, {Uhl}, {Yerushalmi},
  {Smith}, {Liewer}, {Velli}, {Linker}, {Bothmer}, {Rochus}, {Halain}, {Lamy},
  {Auch{\`e}re}, {Harrison}, {Rouillard}, {Patsourakos}, {St. Cyr}, {Gilbert},
  {Maldonado}, {Mariano}, \& {Cerullo}}]{howard2020}
{Howard}, R.~A., {Vourlidas}, A., {Colaninno}, R.~C., {et~al.} 2020, \aap, 642,
  A13, \dodoi{10.1051/0004-6361/201935202}

\bibitem[{{Kasper} {et~al.}(2019){Kasper}, {Bale}, {Belcher}, {Berthomier},
  {Case}, {Chandran}, {Curtis}, {Gallagher}, {Gary}, {Golub}, {Halekas}, {Ho},
  {Horbury}, {Hu}, {Huang}, {Klein}, {Korreck}, {Larson}, {Livi}, {Maruca},
  {Lavraud}, {Louarn}, {Maksimovic}, {Martinovic}, {McGinnis}, {Pogorelov},
  {Richardson}, {Skoug}, {Steinberg}, {Stevens}, {Szabo}, {Velli},
  {Whittlesey}, {Wright}, {Zank}, {MacDowall}, {McComas}, {McNutt}, {Pulupa},
  {Raouafi}, \& {Schwadron}}]{kasper2019}
{Kasper}, J.~C., {Bale}, S.~D., {Belcher}, J.~W., {et~al.} 2019, \nat, 576,
  228, \dodoi{10.1038/s41586-019-1813-z}

\bibitem[{{Kosugi} {et~al.}(2007){Kosugi}, {Matsuzaki}, {Sakao}, {Shimizu},
  {Sone}, {Tachikawa}, {Hashimoto}, {Minesugi}, {Ohnishi}, {Yamada}, {Tsuneta},
  {Hara}, {Ichimoto}, {Suematsu}, {Shimojo}, {Watanabe}, {Shimada}, {Davis},
  {Hill}, {Owens}, {Title}, {Culhane}, {Harra}, {Doschek}, \&
  {Golub}}]{kosugi2007}
{Kosugi}, T., {Matsuzaki}, K., {Sakao}, T., {et~al.} 2007, \solphys, 243, 3,
  \dodoi{10.1007/s11207-007-9014-6}

\bibitem[{{Krieger} {et~al.}(1973){Krieger}, {Timothy}, \&
  {Roelof}}]{krieger1973}
{Krieger}, A.~S., {Timothy}, A.~F., \& {Roelof}, E.~C. 1973, \solphys, 29, 505,
  \dodoi{10.1007/BF00150828}

\bibitem[{{Krucker} {et~al.}(2020){Krucker}, {Hurford}, {Grimm}, {K{\"o}gl},
  {Gr{\"o}belbauer}, {Etesi}, {Casadei}, {Csillaghy}, {Benz}, {Arnold},
  {Molendini}, {Orleanski}, {Schori}, {Xiao}, {Kuhar}, {Hochmuth}, {Felix},
  {Schramka}, {Marcin}, {Kobler}, {Iseli}, {Dreier}, {Wiehl}, {Kleint},
  {Battaglia}, {Lastufka}, {Sathiapal}, {Lapadula}, {Bednarzik}, {Birrer},
  {Stutz}, {Wild}, {Marone}, {Skup}, {Cichocki}, {Ber}, {Rutkowski}, {Bujwan},
  {Juchnikowski}, {Winkler}, {Darmetko}, {Michalska}, {Seweryn}, {Bia{\l}ek},
  {Osica}, {Sylwester}, {Kowalinski}, {{\'S}cis{\l}owski}, {Siarkowski},
  {St{\k{e}}{\'s}licki}, {Mrozek}, {Podg{\'o}rski}, {Meuris}, {Limousin},
  {Gevin}, {Le Mer}, {Brun}, {Strugarek}, {Vilmer}, {Musset}, {Maksimovi{\'c}},
  {F{\'a}rn{\'\i}k}, {Koz{\'a}{\v{c}}ek}, {Ka{\v{s}}parov{\'a}}, {Mann},
  {{\"O}nel}, {Warmuth}, {Rendtel}, {Anderson}, {Bauer}, {Dionies}, {Paschke},
  {Pl{\"u}schke}, {Woche}, {Schuller}, {Veronig}, {Dickson}, {Gallagher},
  {Maloney}, {Bloomfield}, {Piana}, {Massone}, {Benvenuto}, {Massa},
  {Schwartz}, {Dennis}, {van Beek}, {Rodr{\'\i}guez-Pacheco}, \&
  {Lin}}]{krucker2020}
{Krucker}, S., {Hurford}, G.~J., {Grimm}, O., {et~al.} 2020, \aap, 642, A15,
  \dodoi{10.1051/0004-6361/201937362}

\bibitem[{{Larson} {et~al.}(1997){Larson}, {Lin}, {Ergun}, {McTiernan},
  {McFadden}, {Carlson}, {Anderson}, {McCarthy}, {Parks}, {R{\`e}me},
  {Sanderson}, {Kaiser}, \& {Lepping}}]{larson1997}
{Larson}, D.~E., {Lin}, R.~P., {Ergun}, R.~E., {et~al.} 1997, Advances in Space
  Research, 20, 655, \dodoi{10.1016/S0273-1177(97)00453-5}

\bibitem[{{Lemen} {et~al.}(2012){Lemen}, {Title}, {Akin}, {Boerner}, {Chou},
  {Drake}, {Duncan}, {Edwards}, {Friedlaender}, {Heyman}, {Hurlburt}, {Katz},
  {Kushner}, {Levay}, {Lindgren}, {Mathur}, {McFeaters}, {Mitchell}, {Rehse},
  {Schrijver}, {Springer}, {Stern}, {Tarbell}, {Wuelser}, {Wolfson}, {Yanari},
  {Bookbinder}, {Cheimets}, {Caldwell}, {Deluca}, {Gates}, {Golub}, {Park},
  {Podgorski}, {Bush}, {Scherrer}, {Gummin}, {Smith}, {Auker}, {Jerram},
  {Pool}, {Soufli}, {Windt}, {Beardsley}, {Clapp}, {Lang}, \&
  {Waltham}}]{lemen2012}
{Lemen}, J.~R., {Title}, A.~M., {Akin}, D.~J., {et~al.} 2012, \solphys, 275,
  17, \dodoi{10.1007/s11207-011-9776-8}

\bibitem[{{Lockwood}(1995)}]{lockwood1995}
{Lockwood}, M. 1995, \jgr, 100, 21791, \dodoi{10.1029/95JA01340}

\bibitem[{{Lockwood}(1997)}]{lockwood1997}
---. 1997, Annales Geophysicae, 15, 1501, \dodoi{10.1007/s00585-997-1501-4}

\bibitem[{{Lockwood} {et~al.}(1995){Lockwood}, {Davis}, {Smith}, {Onsager}, \&
  {Denig}}]{lockwood1995etal}
{Lockwood}, M., {Davis}, C.~J., {Smith}, M.~F., {Onsager}, T.~G., \& {Denig},
  W.~F. 1995, \jgr, 100, 21803, \dodoi{10.1029/95JA01339}

\bibitem[{{Lockwood} \& {Hapgood}(1998)}]{lockwood1998}
{Lockwood}, M., \& {Hapgood}, M.~A. 1998, \jgr, 103, 26453,
  \dodoi{10.1029/98JA02244}

\bibitem[{{Macneil} {et~al.}(2019){Macneil}, {Owen}, {Baker}, {Brooks},
  {Harra}, {Long}, \& {Wicks}}]{macneil2019}
{Macneil}, A.~R., {Owen}, C.~J., {Baker}, D., {et~al.} 2019, \apj, 887, 146,
  \dodoi{10.3847/1538-4357/ab5586}

\bibitem[{{Maksimovic} {et~al.}(2020){Maksimovic}, {Bale}, {Chust},
  {Khotyaintsev}, {Krasnoselskikh}, {Kretzschmar}, {Plettemeier}, {Rucker},
  {Sou{\v{c}}ek}, {Steller}, {{\v{S}}tver{\'a}k}, {Tr{\'a}vn{\'\i}{\v{c}}ek},
  {Vaivads}, {Chaintreuil}, {Dekkali}, {Alexandrova}, {Astier}, {Barbary},
  {B{\'e}rard}, {Bonnin}, {Boughedada}, {Cecconi}, {Chapron}, {Chariet},
  {Collin}, {de Conchy}, {Dias}, {Gu{\'e}guen}, {Lamy}, {Leray}, {Lion},
  {Malac-Allain}, {Matteini}, {Nguyen}, {Pantellini}, {Parisot}, {Plasson},
  {Thijs}, {Vecchio}, {Fratter}, {Bellouard}, {Lorf{\`e}vre}, {Danto},
  {Julien}, {Guilhem}, {Fiachetti}, {Sanisidro}, {Laffaye}, {Gonzalez},
  {Pontet}, {Qu{\'e}ruel}, {Jannet}, {Fergeau}, {Brochot}, {Cassam-Chenai},
  {Dudok de Wit}, {Timofeeva}, {Vincent}, {Agrapart}, {Delory}, {Turin},
  {Jeandet}, {Leroy}, {Pellion}, {Bouzid}, {Katra}, {Piberne}, {Recart},
  {Santol{\'\i}k}, {Kolma{\v{s}}ov{\'a}}, {Krupa{\v{r}}},
  {Krupa{\v{r}}ov{\'a}}, {P{\'\i}{\v{s}}a}, {Uhl{\'\i}{\v{r}}}, {L{\'a}n},
  {Ba{\v{s}}e}, {Ahl{\`e}n}, {Andr{\'e}}, {Bylander}, {Cripps}, {Cully},
  {Eriksson}, {Jansson}, {Johansson}, {Karlsson}, {Puccio},
  {B{\v{r}}{\'\i}nek}, {{\"O}ttacher}, {Panchenko}, {Berthomier}, {Goetz},
  {Hellinger}, {Horbury}, {Issautier}, {Kontar}, {Krucker}, {Le Contel},
  {Louarn}, {Martinovi{\'c}}, {Owen}, {Retino}, {Rodr{\'\i}guez-Pacheco},
  {Sahraoui}, {Wimmer-Schweingruber}, {Zaslavsky}, \&
  {Zouganelis}}]{maksimovic2020}
{Maksimovic}, M., {Bale}, S.~D., {Chust}, T., {et~al.} 2020, \aap, 642, A12,
  \dodoi{10.1051/0004-6361/201936214}

\bibitem[{{Mandrini} {et~al.}(2014){Mandrini}, {Nuevo}, {V{\'a}squez},
  {D{\'e}moulin}, {van Driel-Gesztelyi}, {Baker}, {Culhane}, {Cristiani}, \&
  {Pick}}]{mandrini2014}
{Mandrini}, C.~H., {Nuevo}, F.~A., {V{\'a}squez}, A.~M., {et~al.} 2014,
  \solphys, 289, 4151, \dodoi{10.1007/s11207-014-0582-y}

\bibitem[{{M{\"u}ller} {et~al.}(2017){M{\"u}ller}, {Nicula}, {Felix},
  {Verstringe}, {Bourgoignie}, {Csillaghy}, {Berghmans}, {Jiggens},
  {Garc{\'\i}a-Ortiz}, {Ireland}, {Zahniy}, \& {Fleck}}]{muller2017}
{M{\"u}ller}, D., {Nicula}, B., {Felix}, S., {et~al.} 2017, \aap, 606, A10,
  \dodoi{10.1051/0004-6361/201730893}

\bibitem[{{M{\"u}ller} {et~al.}(2020){M{\"u}ller}, {St. Cyr}, {Zouganelis},
  {Gilbert}, {Marsden}, {Nieves-Chinchilla}, {Antonucci}, {Auch{\`e}re},
  {Berghmans}, {Horbury}, {Howard}, {Krucker}, {Maksimovic}, {Owen}, {Rochus},
  {Rodriguez-Pacheco}, {Romoli}, {Solanki}, {Bruno}, {Carlsson}, {Fludra},
  {Harra}, {Hassler}, {Livi}, {Louarn}, {Peter}, {Sch{\"u}hle}, {Teriaca}, {del
  Toro Iniesta}, {Wimmer-Schweingruber}, {Marsch}, {Velli}, {De Groof},
  {Walsh}, \& {Williams}}]{muller2020}
{M{\"u}ller}, D., {St. Cyr}, O.~C., {Zouganelis}, I., {et~al.} 2020, \aap, 642,
  A1, \dodoi{10.1051/0004-6361/202038467}

\bibitem[{{Ngampoopun} \& {et al.}(2023)}]{nawin2023}
{Ngampoopun}, N., \& {et al.} 2023, \apj, submitted

\bibitem[{{Owen} {et~al.}(2020){Owen}, {Bruno}, {Livi}, {Louarn}, {Al Janabi},
  {Allegrini}, {Amoros}, {Baruah}, {Barthe}, {Berthomier}, {Bordon},
  {Brockley-Blatt}, {Brysbaert}, {Capuano}, {Collier}, {DeMarco}, {Fedorov},
  {Ford}, {Fortunato}, {Fratter}, {Galvin}, {Hancock}, {Heirtzler}, {Kataria},
  {Kistler}, {Lepri}, {Lewis}, {Loeffler}, {Marty}, {Mathon}, {Mayall}, {Mele},
  {Ogasawara}, {Orlandi}, {Pacros}, {Penou}, {Persyn}, {Petiot}, {Phillips},
  {P{\v{r}}ech}, {Raines}, {Reden}, {Rouillard}, {Rousseau}, {Rubiella},
  {Seran}, {Spencer}, {Thomas}, {Trevino}, {Verscharen}, {Wurz}, {Alapide},
  {Amoruso}, {Andr{\'e}}, {Anekallu}, {Arciuli}, {Arnett}, {Ascolese},
  {Bancroft}, {Bland}, {Brysch}, {Calvanese}, {Castronuovo},
  {{\v{C}}erm{\'a}k}, {Chornay}, {Clemens}, {Coker}, {Collinson}, {D'Amicis},
  {Dandouras}, {Darnley}, {Davies}, {Davison}, {De Los Santos}, {Devoto},
  {Dirks}, {Edlund}, {Fazakerley}, {Ferris}, {Frost}, {Fruit}, {Garat},
  {G{\'e}not}, {Gibson}, {Gilbert}, {de Giosa}, {Gradone}, {Hailey}, {Horbury},
  {Hunt}, {Jacquey}, {Johnson}, {Lavraud}, {Lawrenson}, {Leblanc}, {Lockhart},
  {Maksimovic}, {Malpus}, {Marcucci}, {Mazelle}, {Monti}, {Myers}, {Nguyen},
  {Rodriguez-Pacheco}, {Phillips}, {Popecki}, {Rees}, {Rogacki}, {Ruane},
  {Rust}, {Salatti}, {Sauvaud}, {Stakhiv}, {Stange}, {Stubbs}, {Taylor},
  {Techer}, {Terrier}, {Thibodeaux}, {Urdiales}, {Varsani}, {Walsh}, {Watson},
  {Wheeler}, {Willis}, {Wimmer-Schweingruber}, {Winter}, {Yardley}, \&
  {Zouganelis}}]{owen2020}
{Owen}, C.~J., {Bruno}, R., {Livi}, S., {et~al.} 2020, \aap, 642, A16,
  \dodoi{10.1051/0004-6361/201937259}

\bibitem[{{Owens} {et~al.}(2020){Owens}, {Lockwood}, {Macneil}, \&
  {Stansby}}]{owens2020}
{Owens}, M., {Lockwood}, M., {Macneil}, A., \& {Stansby}, D. 2020, \solphys,
  295, 37, \dodoi{10.1007/s11207-020-01601-7}

\bibitem[{{Owens}(2009)}]{owens2009}
{Owens}, M.~J. 2009, \solphys, 260, 207, \dodoi{10.1007/s11207-009-9442-6}

\bibitem[{{Owens} {et~al.}(2013){Owens}, {Crooker}, \& {Lockwood}}]{owens2013}
{Owens}, M.~J., {Crooker}, N.~U., \& {Lockwood}, M. 2013, Journal of
  Geophysical Research (Space Physics), 118, 1868, \dodoi{10.1002/jgra.50259}

\bibitem[{{Owens} {et~al.}(2018){Owens}, {Lockwood}, {Barnard}, \&
  {MacNeil}}]{owens2018}
{Owens}, M.~J., {Lockwood}, M., {Barnard}, L.~A., \& {MacNeil}, A.~R. 2018,
  \apjl, 868, L14, \dodoi{10.3847/2041-8213/aaee82}

\bibitem[{{Pesnell} {et~al.}(2012){Pesnell}, {Thompson}, \&
  {Chamberlin}}]{pesnell2012}
{Pesnell}, W.~D., {Thompson}, B.~J., \& {Chamberlin}, P.~C. 2012, \solphys,
  275, 3, \dodoi{10.1007/s11207-011-9841-3}

\bibitem[{{Rochus} {et~al.}(2020){Rochus}, {Auch{\`e}re}, {Berghmans}, {Harra},
  {Schmutz}, {Sch{\"u}hle}, {Addison}, {Appourchaux}, {Aznar Cuadrado},
  {Baker}, {Barbay}, {Bates}, {BenMoussa}, {Bergmann}, {Beurthe}, {Borgo},
  {Bonte}, {Bouzit}, {Bradley}, {B{\"u}chel}, {Buchlin}, {B{\"u}chner},
  {Cab{\'e}}, {Cadiergues}, {Chaigneau}, {Chares}, {Choque Cortez}, {Coker},
  {Condamin}, {Coumar}, {Curdt}, {Cutler}, {Davies}, {Davison}, {Defise}, {Del
  Zanna}, {Delmotte}, {Delouille}, {Dolla}, {Dumesnil}, {D{\"u}rig}, {Enge},
  {Fran{\c{c}}ois}, {Fourmond}, {Gillis}, {Giordanengo}, {Gissot}, {Green},
  {Guerreiro}, {Guilbaud}, {Gyo}, {Haberreiter}, {Hafiz}, {Hailey}, {Halain},
  {Hansotte}, {Hecquet}, {Heerlein}, {Hellin}, {Hemsley}, {Hermans}, {Hervier},
  {Hochedez}, {Houbrechts}, {Ihsan}, {Jacques}, {J{\'e}r{\^o}me}, {Jones},
  {Kahle}, {Kennedy}, {Klaproth}, {Kolleck}, {Koller}, {Kotsialos},
  {Kraaikamp}, {Langer}, {Lawrenson}, {Le Clech'}, {Lenaerts}, {Liebecq},
  {Linder}, {Long}, {Mampaey}, {Markiewicz-Innes}, {Marquet}, {Marsch},
  {Matthews}, {Mazy}, {Mazzoli}, {Meining}, {Meltchakov}, {Mercier}, {Meyer},
  {Monecke}, {Monfort}, {Morinaud}, {Moron}, {Mountney}, {M{\"u}ller},
  {Nicula}, {Parenti}, {Peter}, {Pfiffner}, {Philippon}, {Phillips},
  {Plesseria}, {Pylyser}, {Rabecki}, {Ravet-Krill}, {Rebellato}, {Renotte},
  {Rodriguez}, {Roose}, {Rosin}, {Rossi}, {Roth}, {Rouesnel}, {Roulliay},
  {Rousseau}, {Ruane}, {Scanlan}, {Schlatter}, {Seaton}, {Silliman}, {Smit},
  {Smith}, {Solanki}, {Spescha}, {Spencer}, {Stegen}, {Stockman}, {Szwec},
  {Tamiatto}, {Tandy}, {Teriaca}, {Theobald}, {Tychon}, {van Driel-Gesztelyi},
  {Verbeeck}, {Vial}, {Werner}, {West}, {Westwood}, {Wiegelmann}, {Willis},
  {Winter}, {Zerr}, {Zhang}, \& {Zhukov}}]{rochus2020}
{Rochus}, P., {Auch{\`e}re}, F., {Berghmans}, D., {et~al.} 2020, \aap, 642, A8,
  \dodoi{10.1051/0004-6361/201936663}

\bibitem[{{Rodr{\'\i}guez-Pacheco} {et~al.}(2020){Rodr{\'\i}guez-Pacheco},
  {Wimmer-Schweingruber}, {Mason}, {Ho}, {S{\'a}nchez-Prieto}, {Prieto},
  {Mart{\'\i}n}, {Seifert}, {Andrews}, {Kulkarni}, {Panitzsch}, {Boden},
  {B{\"o}ttcher}, {Cernuda}, {Elftmann}, {Espinosa Lara}, {G{\'o}mez-Herrero},
  {Terasa}, {Almena}, {Begley}, {B{\"o}hm}, {Blanco}, {Boogaerts}, {Carrasco},
  {Castillo}, {da Silva Fari{\~n}a}, {de Manuel Gonz{\'a}lez}, {Drews},
  {Dupont}, {Eldrum}, {Gordillo}, {Guti{\'e}rrez}, {Haggerty}, {Hayes},
  {Heber}, {Hill}, {J{\"u}ngling}, {Kerem}, {Knierim}, {K{\"o}hler}, {Kolbe},
  {Kulemzin}, {Lario}, {Lees}, {Liang}, {Mart{\'\i}nez Hell{\'\i}n}, {Meziat},
  {Montalvo}, {Nelson}, {Parra}, {Paspirgilis}, {Ravanbakhsh}, {Richards},
  {Rodr{\'\i}guez-Polo}, {Russu}, {S{\'a}nchez}, {Schlemm}, {Schuster},
  {Seimetz}, {Steinhagen}, {Tammen}, {Tyagi}, {Varela}, {Yedla}, {Yu},
  {Agueda}, {Aran}, {Horbury}, {Klecker}, {Klein}, {Kontar}, {Krucker},
  {Maksimovic}, {Malandraki}, {Owen}, {Pacheco}, {Sanahuja}, {Vainio},
  {Connell}, {Dalla}, {Dr{\"o}ge}, {Gevin}, {Gopalswamy}, {Kartavykh},
  {Kudela}, {Limousin}, {Makela}, {Mann}, {{\"O}nel}, {Posner}, {Ryan},
  {Soucek}, {Hofmeister}, {Vilmer}, {Walsh}, {Wang}, {Wiedenbeck}, {Wirth}, \&
  {Zong}}]{Pacheco2020}
{Rodr{\'\i}guez-Pacheco}, J., {Wimmer-Schweingruber}, R.~F., {Mason}, G.~M.,
  {et~al.} 2020, \aap, 642, A7, \dodoi{10.1051/0004-6361/201935287}

\bibitem[{{Rouillard} {et~al.}(2010{\natexlab{a}}){Rouillard}, {Davies},
  {Lavraud}, {Forsyth}, {Savani}, {Bewsher}, {Brown}, {Sheeley}, {Davis},
  {Harrison}, {Howard}, {Vourlidas}, {Lockwood}, {Crothers}, \&
  {Eyles}}]{rouillard2010a}
{Rouillard}, A.~P., {Davies}, J.~A., {Lavraud}, B., {et~al.}
  2010{\natexlab{a}}, Journal of Geophysical Research (Space Physics), 115,
  A04103, \dodoi{10.1029/2009JA014471}

\bibitem[{{Rouillard} {et~al.}(2010{\natexlab{b}}){Rouillard}, {Lavraud},
  {Davies}, {Savani}, {Burlaga}, {Forsyth}, {Sauvaud}, {Opitz}, {Lockwood},
  {Luhmann}, {Simunac}, {Galvin}, {Davis}, \& {Harrison}}]{rouillard2010b}
{Rouillard}, A.~P., {Lavraud}, B., {Davies}, J.~A., {et~al.}
  2010{\natexlab{b}}, Journal of Geophysical Research (Space Physics), 115,
  A04104, \dodoi{10.1029/2009JA014472}

\bibitem[{{Rouillard} {et~al.}(2011){Rouillard}, {Sheeley}, {Cooper}, {Davies},
  {Lavraud}, {Kilpua}, {Skoug}, {Steinberg}, {Szabo}, {Opitz}, \&
  {Sauvaud}}]{rouillard2011}
{Rouillard}, A.~P., {Sheeley}, N.~R., J., {Cooper}, T.~J., {et~al.} 2011, \apj,
  734, 7, \dodoi{10.1088/0004-637X/734/1/7}

\bibitem[{{Rouillard} {et~al.}(2020{\natexlab{a}}){Rouillard}, {Pinto},
  {Vourlidas}, {De Groof}, {Thompson}, {Bemporad}, {Dolei}, {Indurain},
  {Buchlin}, {Sasso}, {Spadaro}, {Dalmasse}, {Hirzberger}, {Zouganelis},
  {Strugarek}, {Brun}, {Alexandre}, {Berghmans}, {Raouafi}, {Wiegelmann},
  {Pagano}, {Arge}, {Nieves-Chinchilla}, {Lavarra}, {Poirier}, {Amari}, {Aran},
  {Andretta}, {Antonucci}, {Anastasiadis}, {Auch{\`e}re}, {Bellot Rubio},
  {Nicula}, {Bonnin}, {Bouchemit}, {Budnik}, {Caminade}, {Cecconi}, {Carlyle},
  {Cernuda}, {Davila}, {Etesi}, {Espinosa Lara}, {Fedorov}, {Fineschi},
  {Fludra}, {G{\'e}not}, {Georgoulis}, {Gilbert}, {Giunta}, {Gomez-Herrero},
  {Guest}, {Haberreiter}, {Hassler}, {Henney}, {Howard}, {Horbury}, {Janvier},
  {Jones}, {Kozarev}, {Kraaikamp}, {Kouloumvakos}, {Krucker}, {Lagg}, {Linker},
  {Lavraud}, {Louarn}, {Maksimovic}, {Maloney}, {Mann}, {Masson}, {M{\"u}ller},
  {{\"O}nel}, {Osuna}, {Orozco Suarez}, {Owen}, {Papaioannou},
  {P{\'e}rez-Su{\'a}rez}, {Rodriguez-Pacheco}, {Parenti}, {Pariat}, {Peter},
  {Plunkett}, {Pomoell}, {Raines}, {Riethm{\"u}ller}, {Rich}, {Rodriguez},
  {Romoli}, {Sanchez}, {Solanki}, {St Cyr}, {Straus}, {Susino}, {Teriaca}, {del
  Toro Iniesta}, {Ventura}, {Verbeeck}, {Vilmer}, {Warmuth}, {Walsh}, {Watson},
  {Williams}, {Wu}, \& {Zhukov}}]{rouillard2020}
{Rouillard}, A.~P., {Pinto}, R.~F., {Vourlidas}, A., {et~al.}
  2020{\natexlab{a}}, \aap, 642, A2, \dodoi{10.1051/0004-6361/201935305}

\bibitem[{{Rouillard} {et~al.}(2020{\natexlab{b}}){Rouillard}, {Kouloumvakos},
  {Vourlidas}, {Kasper}, {Bale}, {Raouafi}, {Lavraud}, {Howard}, {Stenborg},
  {Stevens}, {Poirier}, {Davies}, {Hess}, {Higginson}, {Lavarra}, {Viall},
  {Korreck}, {Pinto}, {Griton}, {R{\'e}ville}, {Louarn}, {Wu}, {Dalmasse},
  {G{\'e}not}, {Case}, {Whittlesey}, {Larson}, {Halekas}, {Livi}, {Goetz},
  {Harvey}, {MacDowall}, {Malaspina}, {Pulupa}, {Bonnell}, {de Witt}, \&
  {Penou}}]{rouillard2020b}
{Rouillard}, A.~P., {Kouloumvakos}, A., {Vourlidas}, A., {et~al.}
  2020{\natexlab{b}}, \apjs, 246, 37, \dodoi{10.3847/1538-4365/ab579a}

\bibitem[{{Sakao} {et~al.}(2007){Sakao}, {Kano}, {Narukage}, {Kotoku}, {Bando},
  {DeLuca}, {Lundquist}, {Tsuneta}, {Harra}, {Katsukawa}, {Kubo}, {Hara},
  {Matsuzaki}, {Shimojo}, {Bookbinder}, {Golub}, {Korreck}, {Su}, {Shibasaki},
  {Shimizu}, \& {Nakatani}}]{sakao2007}
{Sakao}, T., {Kano}, R., {Narukage}, N., {et~al.} 2007, Science, 318, 1585,
  \dodoi{10.1126/science.1147292}

\bibitem[{{Scherrer} {et~al.}(2012){Scherrer}, {Schou}, {Bush}, {Kosovichev},
  {Bogart}, {Hoeksema}, {Liu}, {Duvall}, {Zhao}, {Title}, {Schrijver},
  {Tarbell}, \& {Tomczyk}}]{scherrer2012}
{Scherrer}, P.~H., {Schou}, J., {Bush}, R.~I., {et~al.} 2012, \solphys, 275,
  207, \dodoi{10.1007/s11207-011-9834-2}

\bibitem[{{Schrijver} \& {De Rosa}(2003)}]{schrijver2003}
{Schrijver}, C.~J., \& {De Rosa}, M.~L. 2003, \solphys, 212, 165,
  \dodoi{10.1023/A:1022908504100}

\bibitem[{{Sheeley} {et~al.}(2009){Sheeley}, {Lee}, {Casto}, {Wang}, \&
  {Rich}}]{sheeley2009}
{Sheeley}, N.~R., J., {Lee}, D.~D.~H., {Casto}, K.~P., {Wang}, Y.~M., \&
  {Rich}, N.~B. 2009, \apj, 694, 1471, \dodoi{10.1088/0004-637X/694/2/1471}

\bibitem[{{Solanki} {et~al.}(2020){Solanki}, {del Toro Iniesta}, {Woch},
  {Gandorfer}, {Hirzberger}, {Alvarez-Herrero}, {Appourchaux}, {Mart{\'\i}nez
  Pillet}, {P{\'e}rez-Grande}, {Sanchis Kilders}, {Schmidt}, {G{\'o}mez Cama},
  {Michalik}, {Deutsch}, {Fernandez-Rico}, {Grauf}, {Gizon}, {Heerlein},
  {Kolleck}, {Lagg}, {Meller}, {M{\"u}ller}, {Sch{\"u}hle}, {Staub}, {Albert},
  {Alvarez Copano}, {Beckmann}, {Bischoff}, {Busse}, {Enge}, {Frahm},
  {Germerott}, {Guerrero}, {L{\"o}ptien}, {Meierdierks}, {Oberdorfer},
  {Papagiannaki}, {Ramanath}, {Schou}, {Werner}, {Yang}, {Zerr}, {Bergmann},
  {Bochmann}, {Heinrichs}, {Meyer}, {Monecke}, {M{\"u}ller}, {Sperling},
  {{\'A}lvarez Garc{\'\i}a}, {Aparicio}, {Balaguer Jim{\'e}nez}, {Bellot
  Rubio}, {Cobos Carracosa}, {Girela}, {Hern{\'a}ndez Exp{\'o}sito}, {Herranz},
  {Labrousse}, {L{\'o}pez Jim{\'e}nez}, {Orozco Su{\'a}rez}, {Ramos},
  {Barandiar{\'a}n}, {Bastide}, {Campuzano}, {Cebollero}, {D{\'a}vila},
  {Fern{\'a}ndez-Medina}, {Garc{\'\i}a Parejo}, {Garranzo-Garc{\'\i}a},
  {Laguna}, {Mart{\'\i}n}, {Navarro}, {N{\'u}{\~n}ez Peral}, {Royo},
  {S{\'a}nchez}, {Silva-L{\'o}pez}, {Vera}, {Villanueva}, {Fourmond}, {de
  Galarreta}, {Bouzit}, {Hervier}, {Le Clec'h}, {Szwec}, {Chaigneau},
  {Buttice}, {Dominguez-Tagle}, {Philippon}, {Boumier}, {Le Cocguen},
  {Baranjuk}, {Bell}, {Berkefeld}, {Baumgartner}, {Heidecke}, {Maue}, {Nakai},
  {Scheiffelen}, {Sigwarth}, {Soltau}, {Volkmer}, {Blanco Rodr{\'\i}guez},
  {Domingo}, {Ferreres Sabater}, {Gasent Blesa}, {Rodr{\'\i}guez
  Mart{\'\i}nez}, {Osorno Caudel}, {Bosch}, {Casas}, {Carmona}, {Herms},
  {Roma}, {Alonso}, {G{\'o}mez-Sanjuan}, {Piqueras}, {Torralbo}, {Fiethe},
  {Guan}, {Lange}, {Michel}, {Bonet}, {Fahmy}, {M{\"u}ller}, \&
  {Zouganelis}}]{solanki2020}
{Solanki}, S.~K., {del Toro Iniesta}, J.~C., {Woch}, J., {et~al.} 2020, \aap,
  642, A11, \dodoi{10.1051/0004-6361/201935325}

\bibitem[{{Spice Consortium} {et~al.}(2020){Spice Consortium}, {Anderson},
  {Appourchaux}, {Auch{\`e}re}, {Aznar Cuadrado}, {Barbay}, {Baudin},
  {Beardsley}, {Bocchialini}, {Borgo}, {Bruzzi}, {Buchlin}, {Burton},
  {B{\"u}chel}, {Caldwell}, {Caminade}, {Carlsson}, {Curdt}, {Davenne},
  {Davila}, {Deforest}, {Del Zanna}, {Drummond}, {Dubau}, {Dumesnil}, {Dunn},
  {Eccleston}, {Fludra}, {Fredvik}, {Gabriel}, {Giunta}, {Gottwald}, {Griffin},
  {Grundy}, {Guest}, {Gyo}, {Haberreiter}, {Hansteen}, {Harrison}, {Hassler},
  {Haugan}, {Howe}, {Janvier}, {Klein}, {Koller}, {Kucera}, {Kouliche},
  {Marsch}, {Marshall}, {Marshall}, {Matthews}, {McQuirk}, {Meining},
  {Mercier}, {Morris}, {Morse}, {Munro}, {Parenti}, {Pastor-Santos}, {Peter},
  {Pfiffner}, {Phelan}, {Philippon}, {Richards}, {Rogers}, {Sawyer},
  {Schlatter}, {Schmutz}, {Sch{\"u}hle}, {Shaughnessy}, {Sidher}, {Solanki},
  {Speight}, {Spescha}, {Szwec}, {Tamiatto}, {Teriaca}, {Thompson}, {Tosh},
  {Tustain}, {Vial}, {Walls}, {Waltham}, {Wimmer-Schweingruber}, {Woodward},
  {Young}, {de Groof}, {Pacros}, {Williams}, \& {M{\"u}ller}}]{spice2020}
{Spice Consortium}, {Anderson}, M., {Appourchaux}, T., {et~al.} 2020, \aap,
  642, A14, \dodoi{10.1051/0004-6361/201935574}

\bibitem[{{Stansby} {et~al.}(2020){Stansby}, {Baker}, {Brooks}, \&
  {Owen}}]{stansby2020}
{Stansby}, D., {Baker}, D., {Brooks}, D.~H., \& {Owen}, C.~J. 2020, \aap, 640,
  A28, \dodoi{10.1051/0004-6361/202038319}

\bibitem[{{Stansby} {et~al.}(2021){Stansby}, {Green}, {van Driel-Gesztelyi}, \&
  {Horbury}}]{stansby2021}
{Stansby}, D., {Green}, L.~M., {van Driel-Gesztelyi}, L., \& {Horbury}, T.~S.
  2021, \solphys, 296, 116, \dodoi{10.1007/s11207-021-01861-x}

\bibitem[{{Stansby} {et~al.}(2019){Stansby}, {Horbury}, \&
  {Matteini}}]{stansby2019}
{Stansby}, D., {Horbury}, T.~S., \& {Matteini}, L. 2019, \mnras, 482, 1706,
  \dodoi{10.1093/mnras/sty2814}

\bibitem[{{The SunPy Community} {et~al.}(2020){The SunPy Community}, Barnes,
  Bobra, Christe, Freij, Hayes, Ireland, Mumford, Perez-Suarez, Ryan, Shih,
  Chanda, Glogowski, Hewett, Hughitt, Hill, Hiware, Inglis, Kirk, Konge, Mason,
  Maloney, Murray, Panda, Park, Pereira, Reardon, Savage, Sipőcz, Stansby,
  Jain, Taylor, Yadav, Rajul, \& Dang}]{sunpy2020}
{The SunPy Community}, Barnes, W.~T., Bobra, M.~G., {et~al.} 2020, The
  Astrophysical Journal, 890, 68, \dodoi{10.3847/1538-4357/ab4f7a}

\bibitem[{{Tian} {et~al.}(2021){Tian}, {Harra}, {Baker}, {Brooks}, \&
  {Xia}}]{tian2021}
{Tian}, H., {Harra}, L., {Baker}, D., {Brooks}, D.~H., \& {Xia}, L. 2021,
  \solphys, 296, 47, \dodoi{10.1007/s11207-021-01792-7}

\bibitem[{{Tsuneta} {et~al.}(2008){Tsuneta}, {Ichimoto}, {Katsukawa}, {Nagata},
  {Otsubo}, {Shimizu}, {Suematsu}, {Nakagiri}, {Noguchi}, {Tarbell}, {Title},
  {Shine}, {Rosenberg}, {Hoffmann}, {Jurcevich}, {Kushner}, {Levay}, {Lites},
  {Elmore}, {Matsushita}, {Kawaguchi}, {Saito}, {Mikami}, {Hill}, \&
  {Owens}}]{tsuneta2008}
{Tsuneta}, S., {Ichimoto}, K., {Katsukawa}, Y., {et~al.} 2008, \solphys, 249,
  167, \dodoi{10.1007/s11207-008-9174-z}

\bibitem[{{van Driel-Gesztelyi} {et~al.}(2012){van Driel-Gesztelyi}, {Culhane},
  {Baker}, {D{\'e}moulin}, {Mandrini}, {DeRosa}, {Rouillard}, {Opitz},
  {Stenborg}, {Vourlidas}, \& {Brooks}}]{vandriel2012}
{van Driel-Gesztelyi}, L., {Culhane}, J.~L., {Baker}, D., {et~al.} 2012,
  \solphys, 281, 237, \dodoi{10.1007/s11207-012-0076-8}

\bibitem[{{Viall} \& {Borovsky}(2020)}]{viall2020}
{Viall}, N.~M., \& {Borovsky}, J.~E. 2020, Journal of Geophysical Research
  (Space Physics), 125, e26005, \dodoi{10.1029/2018JA026005}

\bibitem[{{Walsh} {et~al.}(2020){Walsh}, {Horbury}, {Maksimovic}, {Owen},
  {Rodr{\'\i}guez-Pacheco}, {Wimmer-Schweingruber}, {Zouganelis}, {Anekallu},
  {Bonnin}, {Bruno}, {Carrasco Bl{\'a}zquez}, {Cernuda}, {Chust}, {De Groof},
  {Espinosa Lara}, {Fazakerley}, {Gilbert}, {G{\'o}mez-Herrero}, {Ho},
  {Krucker}, {Lepri}, {Lewis}, {Livi}, {Louarn}, {M{\"u}ller},
  {Nieves-Chinchilla}, {O'Brien}, {Osuna}, {Plasson}, {Raines}, {Rouillard},
  {St Cyr}, {S{\'a}nchez}, {Soucek}, {Varsani}, {Verscharen}, {Watson},
  {Watson}, \& {Williams}}]{walsh2020}
{Walsh}, A.~P., {Horbury}, T.~S., {Maksimovic}, M., {et~al.} 2020, \aap, 642,
  A5, \dodoi{10.1051/0004-6361/201936894}

\bibitem[{{Wang} \& {Ko}(2019)}]{wang2019}
{Wang}, Y.~M., \& {Ko}, Y.~K. 2019, \apj, 880, 146,
  \dodoi{10.3847/1538-4357/ab2add}

\bibitem[{{Wang} {et~al.}(2010){Wang}, {Robbrecht}, {Rouillard}, {Sheeley}, \&
  {Thernisien}}]{wang2010}
{Wang}, Y.~M., {Robbrecht}, E., {Rouillard}, A.~P., {Sheeley}, N.~R., J., \&
  {Thernisien}, A.~F.~R. 2010, \apj, 715, 39,
  \dodoi{10.1088/0004-637X/715/1/39}

\bibitem[{{Wang} \& {Sheeley}(1990)}]{wang1990}
{Wang}, Y.~M., \& {Sheeley}, N.~R., J. 1990, \apj, 355, 726,
  \dodoi{10.1086/168805}

\bibitem[{Wilcox(1968)}]{wilcox1968}
Wilcox, J. 1968, Space Science Reviews, 8, \dodoi{10.1007/BF00227565}

\bibitem[{{Yardley} {et~al.}(2021){Yardley}, {Brooks}, \&
  {Baker}}]{yardley2021}
{Yardley}, S.~L., {Brooks}, D.~H., \& {Baker}, D. 2021, \aap, 650, L10,
  \dodoi{10.1051/0004-6361/202141131}

\bibitem[{{Zirker}(1977)}]{zirker1977}
{Zirker}, J.~B. 1977, Reviews of Geophysics and Space Physics, 15, 257,
  \dodoi{10.1029/RG015i003p00257}

\bibitem[{{Zouganelis} {et~al.}(2020){Zouganelis}, {De Groof}, {Walsh},
  {Williams}, {M{\"u}ller}, {St Cyr}, {Auch{\`e}re}, {Berghmans}, {Fludra},
  {Horbury}, {Howard}, {Krucker}, {Maksimovic}, {Owen},
  {Rodr{\'\i}guez-Pacheco}, {Romoli}, {Solanki}, {Watson}, {Sanchez}, {Lefort},
  {Osuna}, {Gilbert}, {Nieves-Chinchilla}, {Abbo}, {Alexandrova},
  {Anastasiadis}, {Andretta}, {Antonucci}, {Appourchaux}, {Aran}, {Arge},
  {Aulanier}, {Baker}, {Bale}, {Battaglia}, {Bellot Rubio}, {Bemporad},
  {Berthomier}, {Bocchialini}, {Bonnin}, {Brun}, {Bruno}, {Buchlin},
  {B{\"u}chner}, {Bucik}, {Carcaboso}, {Carr}, {Carrasco-Bl{\'a}zquez},
  {Cecconi}, {Cernuda Cangas}, {Chen}, {Chitta}, {Chust}, {Dalmasse},
  {D'Amicis}, {Da Deppo}, {De Marco}, {Dolei}, {Dolla}, {Dudok de Wit}, {van
  Driel-Gesztelyi}, {Eastwood}, {Espinosa Lara}, {Etesi}, {Fedorov},
  {F{\'e}lix-Redondo}, {Fineschi}, {Fleck}, {Fontaine}, {Fox}, {Gandorfer},
  {G{\'e}not}, {Georgoulis}, {Gissot}, {Giunta}, {Gizon}, {G{\'o}mez-Herrero},
  {Gontikakis}, {Graham}, {Green}, {Grundy}, {Haberreiter}, {Harra}, {Hassler},
  {Hirzberger}, {Ho}, {Hurford}, {Innes}, {Issautier}, {James}, {Janitzek},
  {Janvier}, {Jeffrey}, {Jenkins}, {Khotyaintsev}, {Klein}, {Kontar},
  {Kontogiannis}, {Krafft}, {Krasnoselskikh}, {Kretzschmar}, {Labrosse},
  {Lagg}, {Landini}, {Lavraud}, {Leon}, {Lepri}, {Lewis}, {Liewer}, {Linker},
  {Livi}, {Long}, {Louarn}, {Malandraki}, {Maloney}, {Martinez-Pillet},
  {Martinovic}, {Masson}, {Matthews}, {Matteini}, {Meyer-Vernet}, {Moraitis},
  {Morton}, {Musset}, {Nicolaou}, {Nindos}, {O'Brien}, {Orozco Suarez},
  {Owens}, {Pancrazzi}, {Papaioannou}, {Parenti}, {Pariat}, {Patsourakos},
  {Perrone}, {Peter}, {Pinto}, {Plainaki}, {Plettemeier}, {Plunkett}, {Raines},
  {Raouafi}, {Reid}, {Retino}, {Rezeau}, {Rochus}, {Rodriguez},
  {Rodriguez-Garcia}, {Roth}, {Rouillard}, {Sahraoui}, {Sasso}, {Schou},
  {Sch{\"u}hle}, {Sorriso-Valvo}, {Soucek}, {Spadaro}, {Stangalini}, {Stansby},
  {Steller}, {Strugarek}, {{\v{S}}tver{\'a}k}, {Susino}, {Telloni}, {Terasa},
  {Teriaca}, {Toledo-Redondo}, {del Toro Iniesta}, {Tsiropoula}, {Tsounis},
  {Tziotziou}, {Valentini}, {Vaivads}, {Vecchio}, {Velli}, {Verbeeck},
  {Verdini}, {Verscharen}, {Vilmer}, {Vourlidas}, {Wicks},
  {Wimmer-Schweingruber}, {Wiegelmann}, {Young}, \& {Zhukov}}]{zouganelis2020}
{Zouganelis}, I., {De Groof}, A., {Walsh}, A.~P., {et~al.} 2020, \aap, 642, A3,
  \dodoi{10.1051/0004-6361/202038445}

\end{thebibliography}
\bibliographystyle{aasjournal}

\end{document}